\documentstyle[aps,super]{revtex}

\input epsf

\begin{document}

\title{
A symmetric polymer blend confined into a film with antisymmetric surfaces:\\
interplay between wetting behavior and phase diagram
} 

\author{
M.\ M\"{u}ller$^{1,2}$\footnote{email: \tt{Marcus.Mueller@uni-mainz.de}}, E.V.\ Albano$^2$, and K.\ Binder$^1$
\\
${}^1$ {\small Institut f{\"u}r Physik, WA 331, Johannes Gutenberg Universit{\"a}t}
\\
{\small D-55099 Mainz, Germany}
\\
${}^2$ {\small INIFTA, Universidad de La Plata, C.C.\ 16 Suc.\ 4.}
\\
{\small 1900 La Plata, Argentina}
}
\date{\today, draft}
\maketitle

\begin{abstract}
We study the phase behavior of a symmetric binary polymer blend which is confined into a thin
film. The film surfaces interact with the monomers via short range potentials. We calculate the phase behavior
within the self-consistent field theory of Gaussian chains. Over a wide range of parameters
we find strong first order wetting transitions for the semi--infinite system, and the interplay between
the wetting/prewetting behavior and the phase diagram in confined geometry is investigated. Antisymmetric boundaries,
where one surface attracts the $A$ component with the same strength than the opposite surface attracts 
the $B$ component, are applied. The phase transition does not occur close to the bulk critical temperature but in
the vicinity of the wetting transition. For very thin films or weak surface fields one finds a single
critical point at $\phi_c=1/2$. For thicker films or stronger surface fields the phase diagram exhibits 
two critical points and two concomitant coexistence regions.
Only below a triple point there is a single two phase coexistence region. When we 
increase the film thickness the two coexistence regions become the prewetting lines of the semi--infinite system, 
while the triple temperature converges towards the wetting transition temperature from above. The behavior close 
to the tricritical point, which separates phase diagrams with one and two critical points, is studied in the
framework of a Ginzburg--Landau ansatz. Two-dimensional profiles of the interface between the laterally coexisting phases
are calculated, and the interfacial and line tensions analyzed. The effect of fluctuations and corrections to the self-consistent 
field theory are discussed.

\end{abstract}

\section{ Introduction. }
The phase behavior of fluid mixtures in confined geometry has attracted abiding interest over many decades.\cite{EREV,PREV}  The preferential 
interactions at the surfaces give rise to an enrichment of one component at the surface. In a semi-infinite system at phase 
coexistence, the thickness of this enrichment layer diverges at the wetting transition.\cite{CAHN,MICHAEL,DIETRICH,SULIVAN,DEGENNES} Upon approaching the 
wetting transition temperature from below the thickness of the enrichment layer might increase continuously (second order wetting) 
or jump from a microscopically thin layer to a macroscopic layer at the (first order) transition. This latter case is by far the 
most common experimentally. If the transition is of first order a continuation of the singularity persists also slightly above the 
wetting transition temperature. At a chemical potential (partial pressure) of the preferred species, which is smaller than the coexistence value 
(undersaturation), a thin and a thick enrichment layer coexist. Upon following this coexistence line (prewetting) to higher temperatures we decrease the 
difference in the enrichment layers of the coexisting phases and encounter a prewetting critical point.

If a symmetric binary mixture is confined into a film with antisymmetric boundaries, i.e., the upper surface attracts one species with 
exactly the same strength than the lower surface attracts the other species, no phase transition will occur close to the bulk critical point.
Upon decreasing the temperature the enrichment layers at both surfaces gradually develop and stabilize an $AB$ interface in the
center of the film (``soft--mode'' phase). It is only close to the wetting transition temperature that the symmetry is spontaneously broken and the 
$AB$ interface is localized close to one surface. This interface localization--delocalization transition\cite{BROCHARD,PE,SWIFT,BINDER,TROUBLE,GRAVITY} and the anomalous fluctuations of the
delocalized $AB$ interface in the ``soft--mode'' phase have attracted recent interest\cite{BINDER,KLEIN,SF,ANDREAS} and experimental realizations in terms of polymeric 
systems have been investigated.\cite{KLEIN,SF,POLYMER}

The application of experimental techniques (e.g.,  nuclear reaction analysis or neutron reflectometry) is facilitated by the large length scale of 
the enrichment layers, which is determined by the molecules end-to-end distance $R_e$. The macromolecular architecture also allows a successful comparison
to the results of the mean field theory. The free energy cost of an $AB$ interface $\sigma R_e^2$ on the length scale $R_e$ increases with chain length; 
a fact that reduces the effect of interface fluctuation on the phase diagram for very long chains. The extended fractal shape of the polymers leads 
to a strong interdigitation of different molecules. The large number of neighbors with which a molecule interacts strongly suppresses composition 
fluctuations and imparts mean field behavior to the phase diagram except for the ultimate vicinity of the critical point.

In the following we consider a symmetric binary polymer mixture confined into a thin film with antisymmetric boundaries and study how the wetting transition
in the semi-infinite geometry affects the phase stability in a film. We employ self-consistent field calculations\cite{SCF1,SCF2,SCF3,SCF4,SCF5} to calculate the phase diagram
as a function of the incompatibility, the short range surface interactions, and the film thickness. Our paper is arranged as follows: In the
next section we describe the self-consistent field technique.\cite{SCF4,SCF5} Then we present the phase behavior in a thin film with antisymmetric boundaries.
For thick films or strong surface fields the phase diagram 
contains two critical points, corresponding to the prewetting critical points of each surface.
Interfacial profiles between the coexisting, laterally segregated  phases are discussed, and
the interfacial and line tensions are analyzed. The paper closes with a summary and a discussion of fluctuation effects.

\section{Self-consistent field calculations (SCF)}

We consider a binary polymer blend in a volume $V_0=\Delta_0 \times L \times L$. The film contains $n$ polymers.
$\Delta_0$ denotes the film thickness, while $L$ is the lateral extension of the film. Let $\rho$ be the monomer 
number density in the middle of the film. The density at the film surfaces deviates from the density in the middle 
and it is useful to introduce the thickness (volume) $\Delta$ ($V$) of an equivalent film with constant monomer 
density $\Delta \equiv nN/\rho L^2$.

The two surfaces of the film are impenetrable and hard. In a boundary region of width $\Delta_w$ the total 
monomer density drops to zero at both walls. In our calculations we assume the monomer density profile 
$\rho \Phi_0$ to take the form\cite{SCF4,SCF5}
\begin{equation}
\Phi_0(x) = \left\{ \begin{array}{ll}
               \frac{1-\cos\left( \frac{\pi x}{\Delta_w}\right)}{2} & \mbox{for} \qquad 0\leq x \leq \Delta_w \\
	       1                                                    & \mbox{for} \qquad \Delta_w \leq x \leq \Delta_0 - \Delta_w \\
               \frac{1-\cos\left( \frac{\pi (\Delta_0 - x)}{\Delta_w}\right)}{2}  & \mbox{for} \qquad \Delta_0 - \Delta_w \leq x \leq \Delta_0
             \end{array}\right. 
	     \label{eqn:dens}
\end{equation}
A film with the same number of monomers but uniform density would have the thickness $\Delta=\Delta_0-\Delta_w$.
We assume the width of the boundary region to be small compared to the characteristic length scale of the composition
profile, i.e., $\Delta_w/R_e \ll 1$. In accord with previous studies\cite{SCF4,SCF5,PRE1,PRE2} we choose $\Delta_w=0.15 R_e$.
This particular choice of the density profile is employed for computational convenience. If we chose a small value of the ratio
$\Delta_w/R_e$ the results would remain (almost) unaltered but the computational effort (i.e., the required number of basis functions
 (cf.\ below)) would increase substantially.

Both polymer species of the blend -- denoted $A$ and $B$ -- contain the same number of monomeric units $N$ 
and are of the same architecture. We model them as Gaussian chains of end-to-end distance $R_e$.
There is a short range repulsion between the two monomer species which can be parameterized by the 
Flory-Huggins parameter $\chi$. The reduction of the total monomer density imparts also a lower segregation 
to the boundary regions (``missing neighbor'' effect).

Both walls interact with the monomer species via a short range potential. The monomer wall interaction $H$ in units of 
the thermal energy $k_BT$ is modeled as\cite{SCF4,SCF5}:
\begin{equation}
\frac{H(x)}{k_BT} = \left\{ \begin{array}{ll}
            \frac{4 \Lambda_1 R_e}{\Delta_w}\left\{1+\cos\left( \frac{\pi x}{\Delta_w}\right)\right\}                & \mbox{for} \qquad 0\leq x \leq \Delta_w \\
            0                                                                                                           & \mbox{for} \qquad \Delta_w \leq x \leq \Delta_0 - \Delta_w \\
            \frac{4 \Lambda_2 R_e}{\Delta_w}\left\{1+\cos\left( \frac{\pi (\Delta_0 - x)}{\Delta_w}\right)\right\}   & \mbox{for} \qquad \Delta_0 - \Delta_w \leq x \leq \Delta_0
        \end{array}\right.
\end{equation}
A positive value ($H(x)>0$) corresponds to an attraction for the $A$ monomers and a repulsion for the $B$ species.
The range of the monomer wall interaction is assumed to be much smaller than the chain extension and, for convenience,
we employ the same numerical value as for the width of the boundary region in the monomer density profile.
The normalization of the surface fields $\Lambda_1$ and $\Lambda_2$, which act on the monomers close to the 
left and the right wall, is chosen such that the integrated interaction energy between the wall and the 
monomers is independent of the width of the boundary region $\Delta_w$. In the following we consider antisymmetric
surface fields, i.e., $\Lambda \equiv \Lambda_1 = - \Lambda_2$.

The microscopic $A$ monomer density $\hat \Phi_A$ can be expressed as a functional of the polymer 
conformations $\{ {\bf r}_\alpha(\tau)\}$:
\begin{equation}
\hat \Phi_A({\bf r}) = \frac{N}{\rho} \sum_{\alpha=0}^{n_A} \int_0^1 {\rm d}\tau \; \delta\left({\bf r}-{\bf r}_\alpha(\tau)\right)
\end{equation}
where the sum runs over all $n_A$ $A$ polymers in the system and $0 \leq \tau \leq 1$ parameterizes the 
contour of the Gaussian polymer. A similar expression holds for $\hat \Phi_B({\bf r})$.
With this definition the semi-grandcanonical partition function of a binary blend takes the form:
\begin{eqnarray}
{\cal Z} &\sim& \sum_{n_A=1}^n \frac{\exp(+\Delta \mu n_A/2k_BT)}{n_A!}
                              \frac{\exp(-\Delta \mu n_B/2k_BT)}{n_B!} 
             \int {\cal D}_A[{\bf r}]  {\cal P}_A[{\bf r}] 
               \int {\cal D}_B[{\bf r}]  {\cal P}_B[{\bf r}]  \nonumber \\
         &&       \;\times \;\exp\left( - \rho \int {\rm d}^3{\bf r} 
                \left\{ \chi \hat \Phi_A \hat \Phi_B - H({\bf r})(\hat \Phi_A({\bf r})-\hat \Phi_B({\bf r})) \right\}\right)  \nonumber \\
             && \;\times\;\delta\left( \Phi_0({\bf r}) - \hat \Phi_A({\bf r}) - \hat \Phi_B({\bf r}) \right)
\end{eqnarray}
where $n=n_A+n_B$ and $\Delta \mu$ represents the exchange potential between $A$ and $B$ polymers.
The functional integral ${\cal D}$ sums over all chain conformations of the Gaussian polymers and 
${\cal P}[{\bf r}] \sim \exp \left(- \frac{3}{2R_e^2} \int_0^1 {\rm d}\tau \;\left(\frac{{\rm d}{\bf r}}{{\rm d}\tau}\right)^2 \right)$
denotes the statistical weight of a non--interacting Gaussian polymer.
This simple model neglects the coupling between the interaction energy and 
the chain conformations,\cite{CONF} and  a finite stiffness of 
the polymers. Hence, the chain extensions parallel to the walls 
remain always unperturbed.  The Boltzmann factor in the partition function incorporates the thermal 
repulsion between unlike monomers and the interactions between the monomers and the walls. The last factor 
represents the incompressibility of the melt in the center of the film and enforces the monomer density
to decay according to Eq.(\ref{eqn:dens}) at the walls. A finite compressibility of the polymeric fluid,
which results in a reduction of the monomer density at an $AB$ interface, is neglected.

Introducing auxiliary fields $W_A$, $W_B$, $\Phi_A$, $\Phi_B$ and $\Xi$ we rewrite the partition function of the multi--chain system in terms of the partition function of a single chain
\begin{equation}
{\cal Z} \sim \int {\cal D}W_A {\cal D}W_B {\cal D}\Phi_A {\cal D}\Phi_B {\cal D}\Xi \;\;\; \exp \left( -\frac{{\cal G}[W_A,W_B,\Phi_A,\Phi_B,\Xi]}{k_BT}\right)
\label{eqn:Z}
\end{equation}
The free energy functional has the form:

\begin{eqnarray}
\frac{{\cal G}[W_A,W_B,\Phi_A,\phi_B,\Xi]}{n k_BT} & \equiv & +\; \ln\frac{n}{V_0} \
       - \;\ln \Big\{ \exp(\Delta \mu/2k_BT) {\cal Q}_A[W_A] + \exp(-\Delta \mu/2k_BT) {\cal Q}_B[W_B] \Big \} \nonumber \\
    && + \; \frac{1}{V} \int {\rm d}^3{\bf r} \;\; \chi N \Phi_A({\bf r}) \Phi_B({\bf r}) 
       - \; \frac{1}{V} \int {\rm d}^3{\bf r} \;\;  H({\bf r}) N \left\{ \Phi_A({\bf r})-\Phi_B({\bf r})\right\} \nonumber \\
    && - \; \frac{1}{V} \int {\rm d}^3{\bf r} \;\;  \left\{ W_A({\bf r}) \Phi_A({\bf r}) + W_B({\bf r}) \Phi_B({\bf r})\right\} 
       - \; \frac{1}{V} \int {\rm d}^3{\bf r} \;\;  \Xi({\bf r}) \left\{\Phi_0({\bf r}) -  \Phi_A({\bf r}) -  \Phi_B({\bf r})\right\} 
						 \label{eqn:F}
\end{eqnarray}
where ${\cal Q}_A$ denotes the single chain partition in the external field $W_A$:
\begin{equation}
{\cal Q}_A[W_A] = \frac{1}{V_0} \int {\cal D}_1[{\bf r}] {\cal P}_1[{\bf r}] \;\; \exp\left( - \int_0^1 {\rm d}\tau\; W_A({\bf r}(\tau)) \right)
\end{equation}
and a similar expression holds for ${\cal Q}_B$

The functional integration in Eq.(\ref{eqn:Z}) cannot be carried out explicitly. Therefore we employ a 
saddlepoint approximation, which replaces the integral by the largest value of the integrand. This maximum 
occurs at values of the fields and densities determined by extremizing ${\cal G}$ with respect of each of 
its five arguments. These values are denoted by lower--case letters and satisfy the self-consistent set of 
equations:
\begin{eqnarray}
w_A({\bf r}) &=& \chi N \phi_B - H({\bf r}) N + \xi({\bf r}) \nonumber \\
w_B({\bf r}) &=& \chi N \phi_A + H({\bf r}) N + \xi({\bf r}) \nonumber \\
\phi_A({\bf r}) &=& -\frac{V}{{\cal Q}} \frac{{\cal DQ}}{{\cal D}w_A} 
= \frac{V\exp(\Delta \mu/2k_BT)}{V_0{\cal Q}} \int {\cal D}_1{\cal P}_1 \;\; \int_0^1 {\rm d}\tau \;\; \delta({\bf r}-{\bf r}(\tau)) 
                \exp \left( -\int_0^1 {\rm d} \tau \; w_A({\bf r}(\tau))  \right) 
\label{eqn:scf}
\end{eqnarray}
and a similar expression for $\phi_B$. The abbreviation ${\cal Q}$ denotes the semi-grandcanonical single chain partition function
\begin{equation}
{\cal Q} =  \exp(\Delta \mu/2k_BT) {\cal Q}_A + \exp(-\Delta \mu/2k_BT) {\cal Q}_B
\end{equation}

At this stage fluctuations around the most probable configuration are ignored. Most notably, the $AB$ interfaces in the 
self-consistent (SCF) field calculations are ideally flat and there is no broadening by fluctuations of the 
local position of the $AB$ interface (capillary waves).

To calculate the monomer density it is useful to define the end segment distribution $q_A({\bf r},t)$
\begin{equation}
q_A({\bf r},t) = \int_0^t {\cal D}_1[{\bf r}] {\cal P}_1[{\bf r}] \delta({\bf r} - {\bf r} (t)) \exp\left(- \int_0^t {\rm d}\tau\; w_a({\bf r}(\tau)) \right)
\end{equation}
and a similar equation holds for $q_B({\bf r},t)$. The end segment distribution satisfies the diffusion equation:
\begin{equation}
\frac{\partial q_A({\bf r},t)}{\partial t} = \frac{R_e^2}{6} \triangle q_A({\bf r},t) - w_A q_A
\end{equation}
The $A$ monomer density can be expressed via the end segment distribution
\begin{equation}
\phi_A({\bf r}) = \frac{V \exp(\Delta \mu/2k_BT)}{V_0 {\cal Q}}\int_0^1 {\rm d}t\; q_A({\bf r},t) q_A({\bf r},1-t)
\end{equation}
and the single chain partition function is given by:
\begin{equation}
{\cal Q}_A = \frac{1}{V_0} \int {\rm d}^3{\bf r}\; q_A({\bf r} ,1)
\end{equation}
Substituting the saddlepoint values of the densities and fields into the free energy functional (\ref{eqn:F}) we calculate the free energy 
of the different phases.
\begin{equation}
\frac{G}{nk_BT} = - \ln {\cal Q} - \frac{\chi N}{V} \int {\rm d}^3{\bf r}\; \phi_A \phi_B - \frac{1}{V} \int {\rm d}^3{\bf r}\; \xi \Phi_0
\label{eqn:F2}
\end{equation}
The free energy and the monomer densities are invariant under a change $\xi({\bf r}) \to \xi({\bf r}) + c$. Hence, we adjust the constant $c$ 
such that the last term in the equation above vanishes. For an homogeneous bulk (i.e., $\Delta_0 \to \infty$), we obtain from Eq.(\ref{eqn:scf})
\begin{equation}
\frac{\Delta \mu}{k_BT} = \ln \frac{\phi_A}{1-\phi_A} - \chi N (2\phi_A -1)
\label{eqn:mu}
\end{equation}
and Eq.(\ref{eqn:F2}) yields
\begin{equation}
\frac{G}{nk_BT}= \ln \frac{n}{V_0} - \ln 2 {\rm cosh}\left\{ \frac{\chi N}{2} (2\phi_A -1) + \frac{\Delta \mu}{2k_BT}\right\} - \chi N \phi_A(1-\phi_A) + \frac{\chi N}{2}
\end{equation}
The Helmholtz free energy $F$ in the canonical ensemble is related to $G$ via the Legendre transformation
\begin{equation}
F = G + \frac{\Delta \mu}{2} (n_A-n_B)
\end{equation}
and for the homogeneous bulk system $F$ takes the Flory--Huggins form:
\begin{equation}
\frac{F}{nk_BT} = \ln \frac{n}{V_0} +  \phi_A \ln \phi_A + (1-\phi_A) \ln (1-\phi_A) + \chi N \phi_A (1-\phi_A)
\end{equation}

In inhomogeneous systems we expand the spatial dependence of the densities and fields in a set of orthonormal 
functions:
\begin{equation}
f_{kl}(x) = \left\{
\begin{array}{ll}
\sqrt{2} \sin(\pi k x/\Delta_0) & \mbox{for} \; l=0,k=1,2,\cdots \\
\sqrt{2} \sin(\pi k x/\Delta_0) \sqrt{2} \cos (2\pi l y/L)& \mbox{for} \; l>0 , k=1,2,\cdots
\end{array}
\right.
\end{equation}

This procedure results in a set of non--linear equations which are solved by a Newton-Raphson like method. 
For the one--dimensional profiles ($l=0$, $k=1,\cdots$) we use up to 120 basis functions and achieve a relative accuracy $10^{-4}$ in the free energy. 

In order to investigate the interface between laterally coexisting phases we employ two--dimensional SCF calculations in the canonical ensemble. 
In the canonical ensemble the relation between the fields $w_A$ and $w_B$ and the densities is still given by Eq.(\ref{eqn:scf}), but the densities are obtain via:
\begin{equation}
\phi_A({\bf r}) = - \bar \phi_A \frac{V}{{\cal Q}_A} \frac{{\cal DQ}_A}{{\cal D}w_A} 
                = \frac{V \bar \phi_A}{V_0 {\cal Q}_A} \int_0^1 {\rm d}t\; q_A({\bf r},t) q_A({\bf r},1-t)
\end{equation}
$\bar \phi_A$ denotes the average composition of the system. We use 320 basis functions for film thickness $\Delta_0=0.9 R_e$. 
The Helmholtz free energy takes the form:
\begin{equation}
\frac{F}{nk_BT} = \bar \phi_A \ln \bar \phi_A + (1-\bar \phi_A) \ln (1-\bar \phi_A) - \bar \phi_A \ln {\cal Q}_A - (1-\bar \phi_A) \ln {\cal Q}_B 
- \frac{\chi N}{V}  \int {\rm d}^3{\bf r}\; \phi_A\phi_B
\end{equation}
where we have used $\int {\rm d}^3{\bf r}\; \xi \Phi_0=0$.

The temperature scale in the SCF calculations is set by the incompatibility $\chi N$, the length
scale is set by the molecule's end-to-end distance $R_e$, and the strength of the surface fields
appear only in the combination $\Lambda_1N$ and $\Lambda_2N$. Moreover, we employ the reduced chain 
length $\bar{N} = (\rho R_e^3/N)^2$ to measure the degree of mutual interdigitation. In the framework of the 
SCFT, systems with the same values of $\chi N$ and $R_e$ but different $\bar N$ exhibit identical behavior.
As we shall discuss, the mean field approximation is appropriate in many aspects in the limit $\bar{N} \to \infty$,
whereas there are corrections to the SCF calculations for finite $\bar N$. A different interesting behavior emerges 
at strong segregation $\chi N \to \infty$ (SSL). In this regime, many properties of the SCF calculations are describable by 
simple analytical expressions, and we shall denote these expressions by SSL in the following.

\section{ Results. }
\subsection{Bulk phase diagram and wetting behavior.}

Coexistence between different phases occurs if the two phases have the same semi-grandcanonical energy at fixed temperature $1/\chi N$ and exchange
potential $\Delta \mu$. Since the bulk is symmetric with respect to exchanging $A \rightleftharpoons B$, phase coexistence occurs at $\Delta \mu_{\rm coex}=0$ 
and the phase diagram is given implicitly by Eq.(\ref{eqn:mu}). The critical temperature is given by $1/\chi_cN = 1/2$. Of course, the SCF theory yields
a parabolic shape of the binodal close to the critical point, because it is a mean--field theory.

The location of the wetting transition can be determined via the Young equation.\cite{YOUNG} The $A$ component wets the surface, if
$\sigma_{WB} - \sigma_{WA} > \sigma_{AB}$,
where $\sigma_{AB}$ denotes the $AB$ interface tension between the coexisting bulk phases, and $\sigma_{WA}$ and $\sigma_{WB}$ denote the excess surface free energies per 
unit area of a surface in contact with the $A$--rich or $B$--rich phase, respectively. In the strong segregation limit (SSL), i.e., $\chi N \gg 2$, the $AB$ interface tension 
takes the form:\cite{SEM}
\begin{equation}
\frac{\sigma_{AB}R_e^2}{k_BT}  = \sqrt{\bar{N}} \sqrt{\frac{\chi N}{6}} \left(1-\frac{4\ln 2}{\chi N} + \cdots\right) \qquad \mbox{(SSL)}
\end{equation}
The excess surface free energy of a surface in contact with the $A$--rich phase has two contributions. On the one hand the polymer conformations are restricted
due to the presence of the surface and the decay of the density profile in the vicinity of the wall. Since the $A$ and $B$ polymers have identical architecture this
conformational entropy contribution to the excess surface free energy is, however, the same for the two species and does not enter into the difference 
$\sigma_{WB}-\sigma_{WA}$.\cite{WET} On the other hand, the surface fields $\Lambda_1$ and $\Lambda_2$ give rise to a contribution to the excess surface free energy.
If the surface is completely covered by the $A$ component, this contribution amounts to:
\begin{equation}
\frac{E_w}{L^2 k_BT} = \rho \int_0^{\Delta_0/2} {\rm d}x\; H(x) \Phi_0(x) = \rho R_e \Lambda
\label{eqn:Ewall}
\end{equation}
The contribution of a surface covered by the $B$ component has the opposite sign. 

If the wetting transition occurs at high incompatibility it will be of first order. In this case
the enrichment layer in the non--wet state is negligible small. Monte Carlo simulations show that
this is a good approximation.\cite{WET} The Young equation\cite{YOUNG} yields for the strength of the surface field at the wetting transition:
\begin{equation}
\Lambda_{{\rm wet}} N \approx \sqrt{\frac{\chi_{\rm wet} N}{24}}  \left(1-\frac{4\ln 2}{\chi_{\rm wet} N} + \cdots\right) \qquad \mbox{(SSL, first order wetting)}
\label{eqn:wetssl}
\end{equation}
If the integrated monomer-wall interaction $\Lambda R_e \sim \Lambda \sqrt{N}$ -- which is the experimentally relevant quantity --
does not depend on the chain length $N$, the left hand side of the equation will be large and the wetting transition in the binary 
polymer blend will occur in the strong segregation limit, i.e., $\chi_{\rm wet} \sim (\Lambda R_e)^2 \gg 1/N$. This is in contrast to the behavior 
of mixtures of small molecules, where the Cahn argument\cite{CAHN} suggests that the wetting transition 
occurs close to the critical point. 

If the wetting transition is first order a prewetting lines emanates from the coexistence curve
above the wetting transition temperature. Along this line a thin and a thick enrichment layer
coexist at undersaturation. For short range forces the prewetting line approaches the bulk
coexistence curve linearly $\Delta \mu_{\rm prewet} \sim (T-T_{\rm wet})/\ln(T-T_{\rm wet})$.\cite{HS}
Upon increasing the temperature, the difference in the thickness of the enrichment layers
decreases and the prewetting line ends in a prewetting critical point. This prewetting behavior
is pertinent to the phase behavior in thin films.

Only for very weak surface fields the wetting transition occurs
close to the critical point. In this limit polymers exhibit a behavior similar to small molecules
and the wetting behavior has been studied within the  square--gradient approximation.\cite{SG1,SG2} The latter assumes
that the concentration varies slowly on the length scale $R_e$. In this approximation the dependence
of the bare surface free energy on the composition at the wall plays a central role. In our model, both 
the surface fields and the ``missing neighbor'' effect due to the decay of the monomer density at the wall
give rise to a composition dependence of the bare surface free energy.
\begin{eqnarray}
f^{\rm bare}_{\rm wall}(\phi_{As}) &=& \frac{\Delta F}{\rho L^2k_BT} =  - \mu_1 \phi_{As} - \frac{1}{2} g_1 \phi_{As}^2 \nonumber \\
                                &=& \int_0^{\Delta_w} {\rm d}z\; \left\{ (\phi_A-\phi_B) H + \chi \phi_A\phi_B\right\} - \chi \phi_{As}\phi_{Bs} \frac{\Delta_w}{2} \nonumber \\
				&=& - (\phi_{As} -\phi_{Bs}) R_e \Lambda - \frac{1}{8} \Delta_w \chi \phi_{As}\phi_{Bs}  
\end{eqnarray}
where $\phi_{As}=\lim_{z \to 0} \phi_A(z)/\Phi_0(z)$ denotes the composition at the surface.
Other contributions to the bare surface free energy (e.g., terms proportional to the gradient of the composition at the surface) are omitted.
From this we identify the coefficients $\mu_1=2 \Lambda R_e + \chi \Delta_w/8$ and $g_1= - \chi \Delta_w/4$. 
One central result of the square gradient theory is that wetting transitions close to the critical point are of second order 
and occur at $ \mu_1= - g_1(1-\phi_{A}^{ \rm bulk})$ 
(a detailed derivation of this equation in the framework of the square gradient approximation can be found in Ref\cite{SG1}).
Using the parameters of our model we rewrite this result in the form:
\begin{equation}
\Lambda_{\rm wet} N \approx \frac{1}{16} \chi_{\rm wet} N \frac{\Delta_w}{R_e} ( 1- 2\phi_{A}^{\rm bulk}) \qquad \mbox{(WSL, second order wetting)}
\label{eqn:2nd}
\end{equation}
For arbitrary strength of the surface fields we expect the variable $\Lambda_{{\rm wet}} N$ to be a function of $\chi_{\rm wet} N$ and the above
equations describe the limit $\chi N \to \infty$ and $\chi N \to 2$.

\subsection{Interface localization--delocalization transition.}
Rather than focusing on the detailed composition profile across the enrichment layers at the surface much qualitative insight 
into the wetting behavior and the interface localization--delocalization transition can be gained from characterizing the profile 
only by the distance $l$ between the wall and the $AB$ interface. The dependence of the free energy per unit area 
-- the effective $AB$ interface potential $g_{\rm wall}(l)= F(l)/L^2=\rho k_BT \Delta f/N$ -- on the distance $l$ determines 
the wetting behavior.  
The short range surface fields distort the profile in the vicinity of the wall and give rise to an effective interaction which
decays exponentially with the distance $l$ between the wall and the $AB$ interface. Qualitatively, the effective 
interface potential $g_{\rm wall}(l)$ in the semi--infinite system can be expanded in the form\cite{MICHAEL}
\begin{equation}
g_{\rm wall}(l) = a(\chi N) \exp(-\lambda l) - b \exp(-2\lambda l) + c \exp(-3 \lambda l) \label{eqn:eih}
\end{equation}
This expression retains only the  lowest powers of $\exp(-\lambda l)$,
which are necessary to bring about the salient features of the wetting behavior. We neglect the temperature dependence of the coefficients 
$b$ and $c$. $\lambda$ denotes the length scale of the interaction between the surface and the $AB$ interface (cf.\ below). 
In principle, the numerical values of the coefficients can be obtained from fitting the results of our SCF calculation to the equation above,
but we cannot offer analytical expression for the coefficients in the framework of the SCFT. In the following we discuss the qualitative behavior 
which arises from an effective interface potential of type (\ref{eqn:eih}).

\begin{figure}[htbp]
    \begin{minipage}[t]{160mm}%
       \mbox{
        ({\bf a})
        \setlength{\epsfxsize}{8cm}
        \epsffile{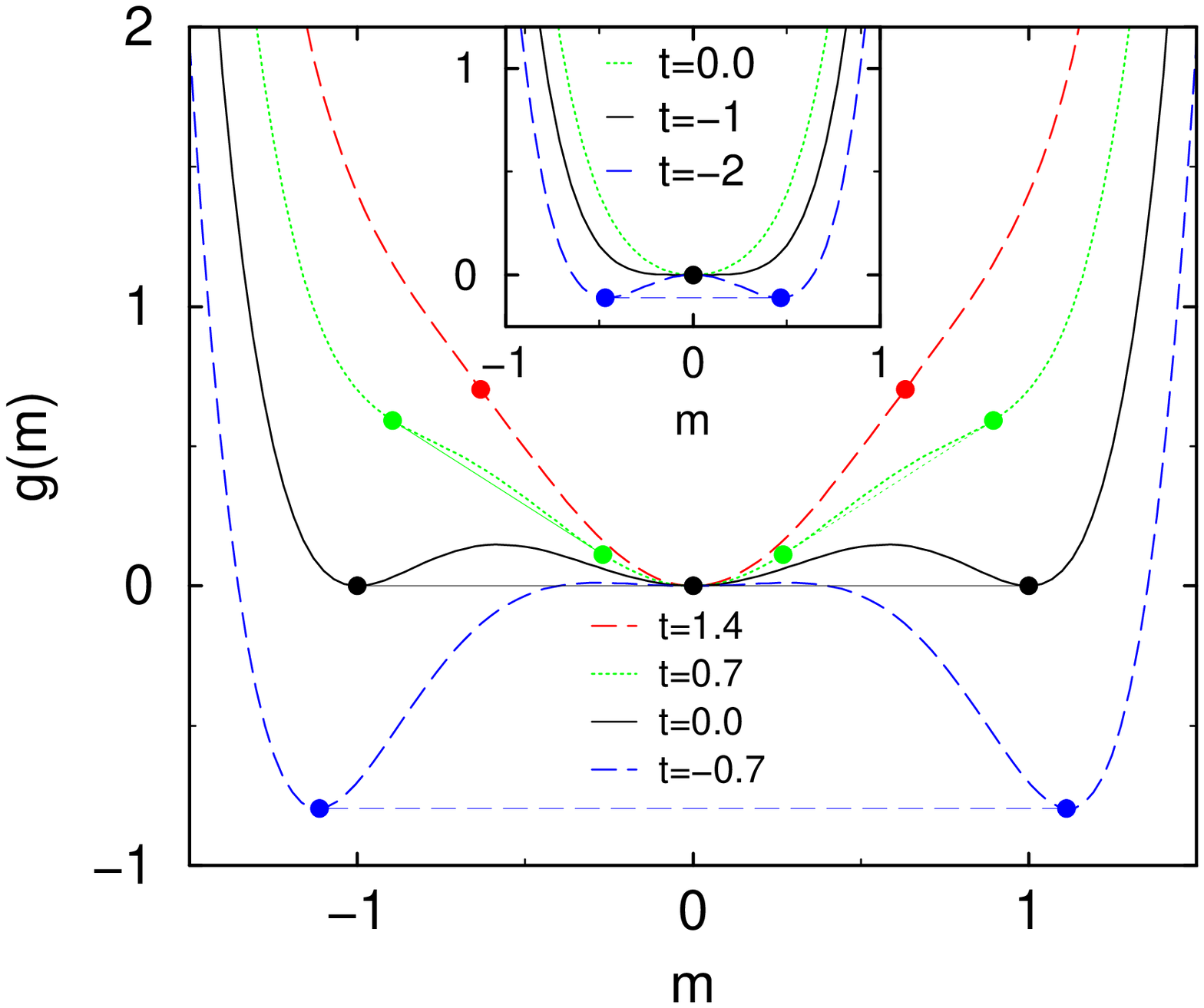}
        ({\bf b})
        \setlength{\epsfxsize}{8.504cm}
        \epsffile{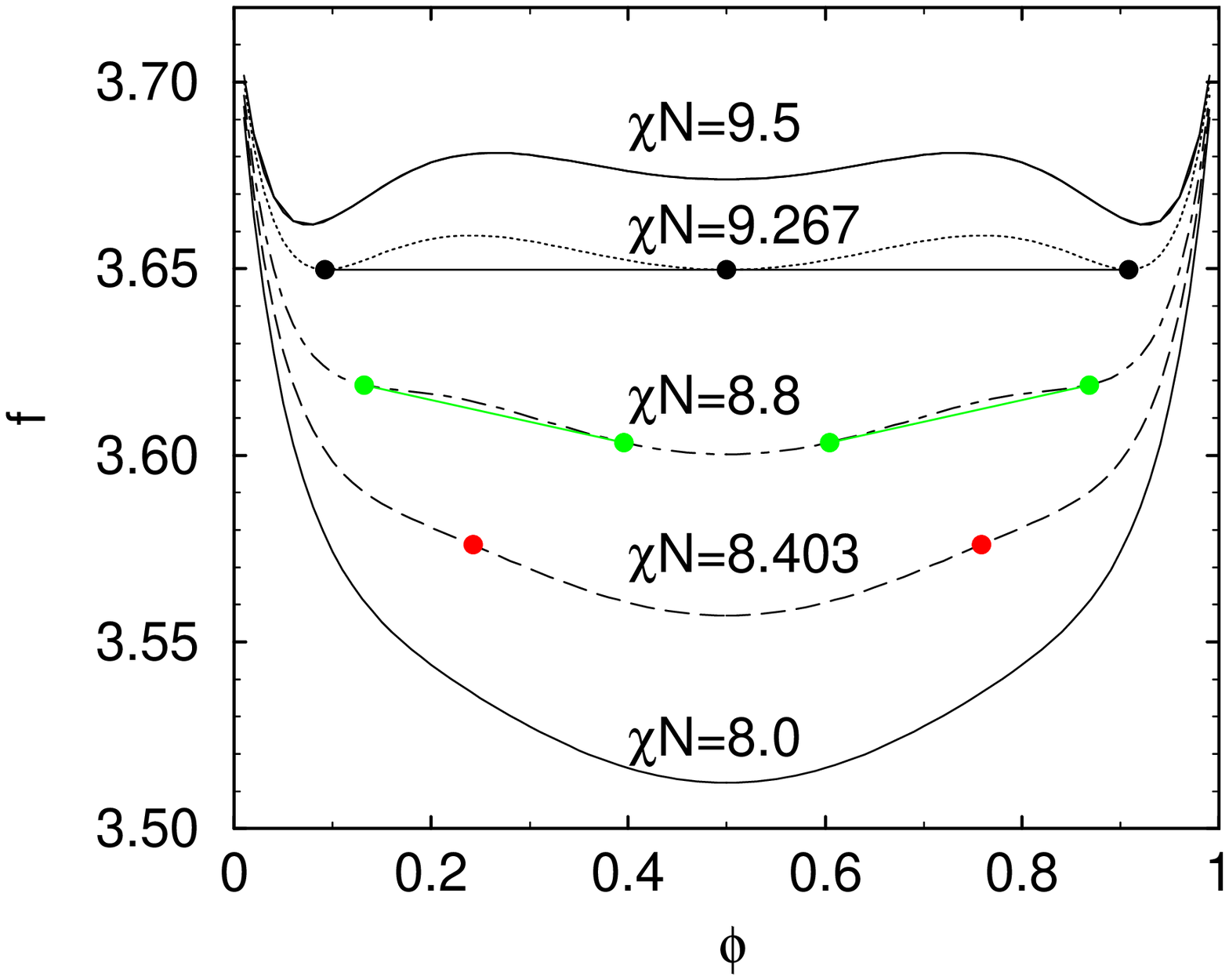}
       }
    \end{minipage}%
    \hfill%
    \begin{minipage}[b]{160mm}%
    \vspace*{1cm}
       \caption{
       \label{fig:ginz} 
       ({\bf a})
       Schematical illustration of the Ginzburg--Landau free energy for the case of the two critical 
       points ($r=1>0$) and for a single second order transition ($r=-1<0$) (inset).
       The values of the temperature--like variable $t$ are given in the key. Coexisting values and 
       critical points are marked by points, while straight lines present
       Maxwell constructions connecting the two coexisting phases.
       ({\bf b})
       free energy in the canonical ensemble as a function of \protect$\chi N$ and \protect$\phi$
       for \protect{$\Delta_0=0.9$} and \protect{$\Lambda N =0.5$}. The temperatures correspond to a supercritical isotherm
       the critical isotherm, the triple temperature and an even lower temperature.
       }
    \end{minipage}%
\end{figure}

If the coefficient $b$ is negative the wetting
transition is of second order and occurs when the coefficient $a$ changes its sign. Following Parry and Evans\cite{PREV,PE} we obtain the effective 
interface potential $g(l)$ in a thin film by superimposing the interactions originating from each individual wall;
$g(l)=g_{\rm wall}(l)+g_{\rm wall}(\Delta-l)$.
The qualitative form of the potential is shown in the inset of Fig.\ref{fig:ginz}({\bf a}), where the parameter 
$m \sim l-\Delta/2$ is proportional to the distance of the interface from the center of the film. The values
of $t$ correspond to various temperatures.
A second order wetting transition gives rise to a second order
interface localization--delocalization transition. Above the transition a single $AB$ interface parallel to the surfaces
is stable, the system is in the one phase region. Below the transition, the system phase separates laterally into a 
phases where the $AB$ interface is located close to the right or the left surface, respectively. The transition is of second 
order, i.e., the composition difference between the coexisting phases increases continously. The transition temperature
approaches rapidly the wetting transition temperature of the semi--infinite system from below as we increase the film thickness.

If the coefficient $b$ is positive, the form of $g_{\rm wall}(l)$ leads to a first order wetting transition in the semi--infinite system
where a thin layer of thickness $l_-=1/\lambda \ln (2c/b)$ coexists with a macroscopically thick enrichment layer at $a_{\rm wet}=b^2/4c$. 
The prewetting critical point is located at $a_{\rm pwc}= 16 a_{\rm wet}/9$ and $l_{\rm pwc}=1/\lambda \ln (9c/2b)$. 
The superposition of interactions between the $AB$ interface and the opposing walls yields the effective interface potential
in a thin film. The qualitative shape of the potential is shown in Fig.\ref{fig:ginz} ({\bf a}) schematically,
while panel ({\bf b}) presents the results of the SCF calculations for $\Delta_0=0.9 R_e$and $\Lambda N=0.5$.

At low temperature we find phase coexistence between two laterally segregated phases. Upon increasing the temperature we encounter
a triple point at which these two phases, where the $AB$ interface is located close to the right or the left surface, respectively, coexist
with third phase, where the $AB$ interface is delocalized at  the center of the film. The location of the triple point ($l_{\rm triple}$ and $a_{\rm triple}$) 
is given by the conditions $g(l_t)=g(\Delta/2)$ and $\partial g/\partial l|_{l_t}=0$. For large film thickness this yields: 
$a_{\rm triple} - a_{\rm wet} = b \exp(-\lambda D/2) + {\cal O}(\exp(-\lambda D))$; i.e., as the film thickness is increased the triple 
temperature converges exponentially fast to the temperature of the first order wetting transition of the semi--infinite system from above.
Above this triple point there are two coexistence regions which each correspond to the prewetting coexistence of the semi--infinite system.
At $\Delta \mu<0$ we find the coexistence of a thick and a thin enrichment layer of the $A$ species at the $A$--attracting surface and at $\Delta \mu>0$ a
similar coexistence at the opposite surface. The two coexistence regions end in critical points close to the prewetting critical temperature 
of the semi--infinite system. This first order interface localization--delocalization behavior is the analogon of the first order wetting behavior
of the semi--infinite system.

\begin{figure}[htbp]
    \begin{minipage}[t]{160mm}%
       \mbox{
        \setlength{\epsfxsize}{8cm}
        \epsffile{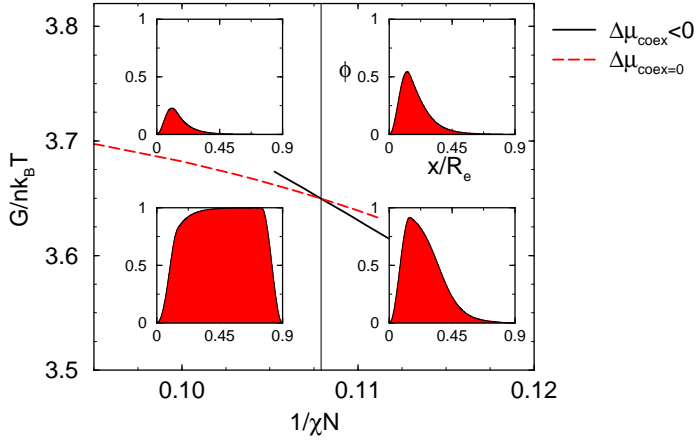}
       }
    \end{minipage}%
    \hfill%
    \begin{minipage}[b]{160mm}%
    \vspace*{1cm}
       \caption{
       \label{fig:gfree} 
       \protect{
       Semi-grandcanonical free energy $G$ of the coexisting phases at $\Lambda N =0.5$
       and $\Delta_0= 0.9 R_e$. For $\Delta \mu_{\rm coex} \equiv 0$ and $T<T_t=0.108$ an $A$-rich phase 
       coexists with a $B$-rich phase, while for $\Delta \mu_{\rm coex}<0$ and $T>T_t$ both 
       coexisting phases are $B$-rich and differ in the thickness of the $A$ layer at the surface.
       For $T>T_t$ there exists another coexistence region with $\Delta \mu_{\rm coex}>0$ which is related
       to one displayed by exchanging $A$ vs.\ $B$.
       The insets show the $A$ density profiles across the film for the coexisting 
       phases at $\chi N = 10$ (left) and $\chi N = 8.7$ (right).
       }
       }
    \end{minipage}%
\end{figure}

The different coexisting phases and their semi--grandcanonical free energy $G$ are presented in Fig.\ref{fig:gfree} for $\Delta_0=0.9 R_e$
and $\Lambda N =0.5$. Below the triple point $1/\chi N < 0.108$ the phases are well segregated. The monomer density profiles of the $A$--component
are depicted on the left side. Upon following the coexistence curve to higher temperatures $G$ decreases. At the triple temperature, these two phases
coexist with a third phase in which the interface is delocalized in the middle of the film. From there onwards, there are two coexistence regions at
positive and negative values of the exchange potential $\Delta \mu$. Profiles of the two phases of the $A$--poor coexisting region are presented 
on the right side. They consist of a thin (upper right inset of Fig.\ref{fig:gfree}) and a thin (lower right inset) enrichment layer of the $A$ component at the surface that
favors $A$.

For our strictly antisymmetric system, the concentration corresponding 
to the triple point always is exactly $1/2$ due to the symmetry. This has an interesting consequence if 
one cools a mixture at $\phi=1/2$: while in the bulk this mixture would undergo a second order phase separation 
(critical unmixing at $\chi=\chi_{\rm crit}=2/N$, $\phi=\phi_{\rm crit}=1/2$), one finds a single first order 
unmixing transition at $\chi=\chi_{\rm triple}$. For asymmetric compositions, however, enrichment layers form
gradually at a wall close to the bulk critical temperature. This stabilizes an $AB$ interface, which runs parallel 
to the surfaces. The interface is located close to one surface; its position is given by the composition of the system.
Close to the prewetting critical point, the enrichment layer may phase separate laterally into a thick and a thin enrichment 
layer. Upon further cooling, we encounter a second phase transition where the thickness of the thick enrichment layer 
become comparable to the film thickness, i.e., two almost completely segregated phases coexist.

\begin{figure}[htbp]
    \begin{minipage}[t]{160mm}%
       \mbox{
        \setlength{\epsfxsize}{8cm}
        \epsffile{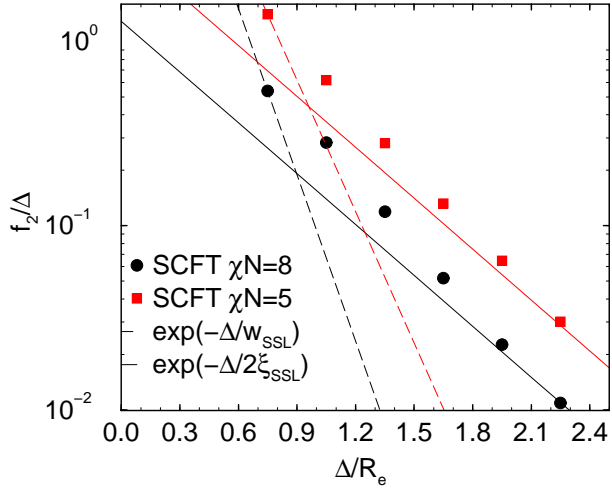}
       }
    \end{minipage}%
    \hfill%
    \begin{minipage}[b]{160mm}%
    \vspace*{1cm}
       \caption{
       \label{fig:range}
       Curvature of the interface potential at the center of the film. Symbols
       represent SCF calculations for \protect$\chi N=5$ and $8$, and  \protect$\Lambda N = 0.5$.
       Solid lines correspond to \protect$g \sim f_2/\Delta \sim \exp(-\lambda \Delta/2)$
       with \protect$1/\lambda = \xi$, while dashed lines depict the behavior for \protect$1/\lambda = w/2$.
       }
    \end{minipage}%
\end{figure}

Previous Monte Carlo simulations\cite{WET} yield evidence that the interaction range $1/\lambda$ in Eq.(\ref{eqn:eih}) is determined by the bulk correlation length 
$\xi$ for large distances between the $AB$ interface and the surface. This is in accord with the expectation that the $AB$ interface 
profile in the outer wing is characterized by the length scale $\xi$, which measures the decay of composition fluctuations in the bulk, 
rather than $w/2$, which characterizes the slope of the $AB$ interface profile at the center of the interface. This is further corroborated 
by our SCF calculations. For two temperatures $\chi N =5$ and $8$ above the critical temperature we have
measured the free energy density $f$ as a function of the composition $\phi$ for various film thicknesses. Around $\phi=1/2$ the
free energy density can be expanded in the form $f = f_1 + f_2(\phi-1/2)^2$. This yields for the effective $AB$ interface potential 
$g(l)= \rho k_BT \Delta f/N \sim {\rm const} + f_2(l-\Delta/2)^2/\Delta$. Above the critical temperature we can estimate the effective
range $1/\lambda$ of the interaction according to $g(l) \sim \exp(-\lambda \Delta/2) \sim f_2/\Delta$. In Fig.\ref{fig:range} we plot
$f_2/\Delta$ {\em vs.}\ the film thickness $\Delta$. For large film thicknesses the data exhibit an exponential dependence on the film thickness.
Upon increasing the temperature the interaction increases as the surfaces repel the $AB$ interface stronger. For large $\Delta$ the
interaction range is compatible with $1/\lambda = \xi \approx R_e/\sqrt{18}$, where we have used the behavior of the correlation length at strong segregation.
For small $\Delta$, i.e., distances between the $AB$ interface and the surface which are not very much larger than the interfacial width $w$, the interaction 
decays somewhat faster $w < 1/\lambda < \xi$. At these intermediate distances a rather complicated interaction has been predicted.\cite{SEM}

Upon varying the sign of the coefficient $b$ we alter the order of the interface localization--delocalization transition. 
At the tricritical point the order of the transition changes. For small values of $b$ we make a phenomenological Ginzburg-Landau 
ansatz for the effective interface potential $g(m)$ in terms of the (not normalized) order parameter  $m \sim  \phi-1/2 \sim l-\Delta/2$. 
For antisymmetric surface fields the  effective interface potential is invariant under the transformations $A \rightleftharpoons B$ and 
must be an even function of $m$. We assume the simplest ansatz which allows for three phase coexistence:
\begin{equation}
g(m) = m^2 \left(m^2-r\right)^2 + t m^2
\label{eqn:landau}
\end{equation}
The coefficients of this Landau expansion (\ref{eqn:landau}) can be derived from an effective interface Hamiltonian (\ref{eqn:eih}):
$r = -15 g_4/g_6$ and $t=360 g_2/g_6 - 225 g_4^2/g_6^2$, where $g_n \equiv \partial^n g_{\rm wall}/\partial l^n|_{\Delta/2}$
denotes the $n$th derivative of the wall--interface potential at the center of the film.

This effective interface potential is depicted in Fig.\ref{fig:ginz}({\bf a}).  For $r<0$ (inset) the coefficient in front of the fourth order term $m^4$ is positive and we find 
a single second order phase transition at $t_c = - r^2$, $m_c=0$, and $\mu_c=0$.
The case $r>0$ corresponds to a first order interface localization--delocalization transition.  For $t=0$ there is a 
three phase coexistence, at which the order parameter of the coexisting phases takes the values $0$ and $\pm \sqrt{r}$. 
Of course, the Landau expansion is only appropriate for small $r$. The parameter $t$ characterizes the temperature difference 
to this triple point. Above the triple temperature, we find two coexistence regions and, eventually, we encounter two critical points 
at $t_c = 7 r^2/5$ and order parameters $m_c = \pm \sqrt{2r/5}$. The critical chemical potential $\mu_c = \partial f/\partial m|_c$ is 
given by $64\sqrt{2}r^{5/2}/25\sqrt{5}$.
$r=0$ marks the tricritical transition; the three coexisting phases collapse to a single one with order parameter $m=0$. 
The critical and triple temperature coincide likewise. The fourth order coefficient in the Ginzburg--Landau ansatz vanishes and the binodals close to the critical 
temperature open like $m \sim \pm (-t)^\beta$ with $\beta_{\rm tri}=1/4$ rather than with $\beta=1/2$ (in mean field approximation).

\begin{figure}[htbp]
    \begin{minipage}[t]{160mm}%
       \mbox{
        ({\bf a})
        \setlength{\epsfxsize}{8cm}
        \epsffile{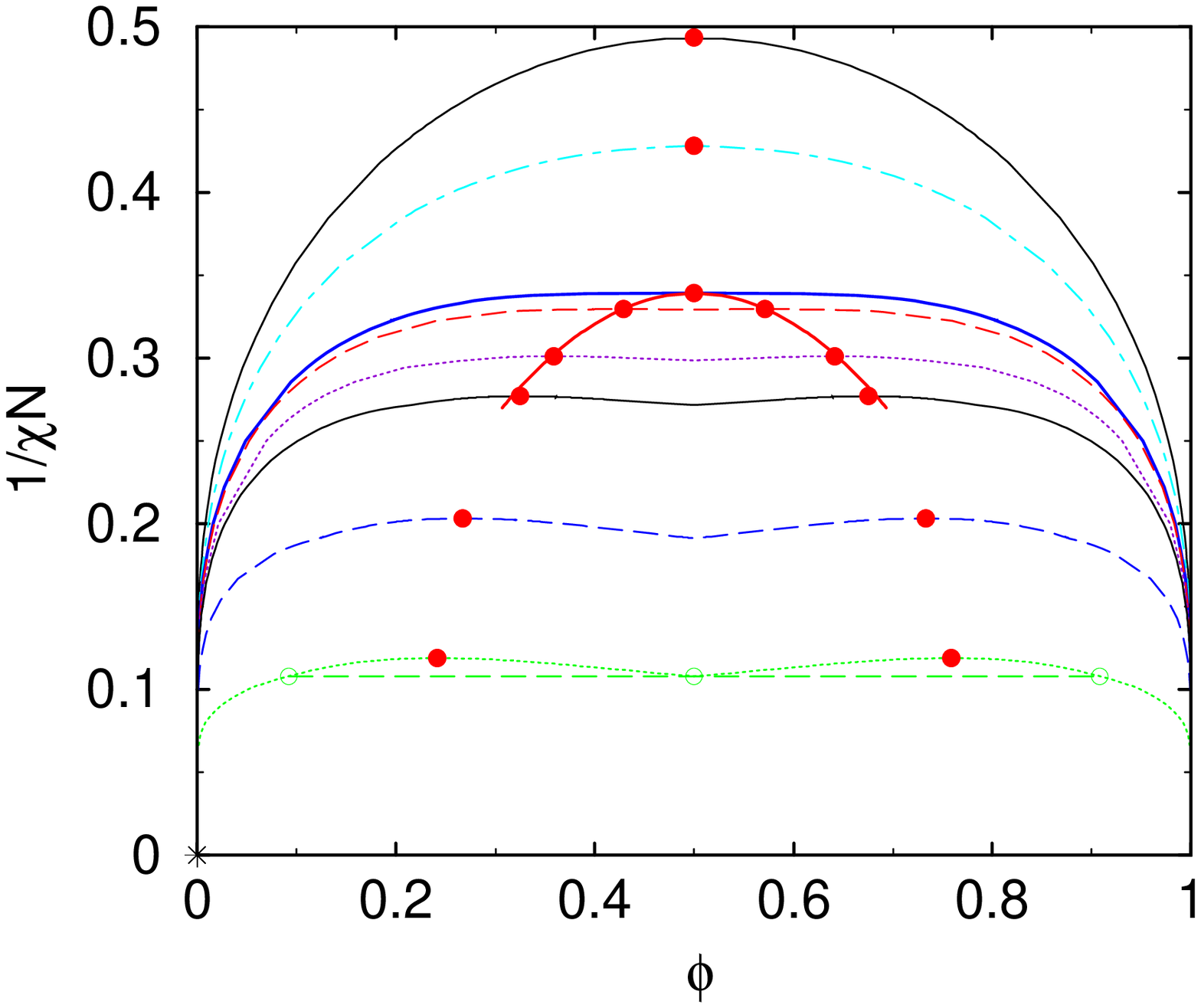}
        ({\bf b})
        \setlength{\epsfxsize}{8.224cm}
        \epsffile{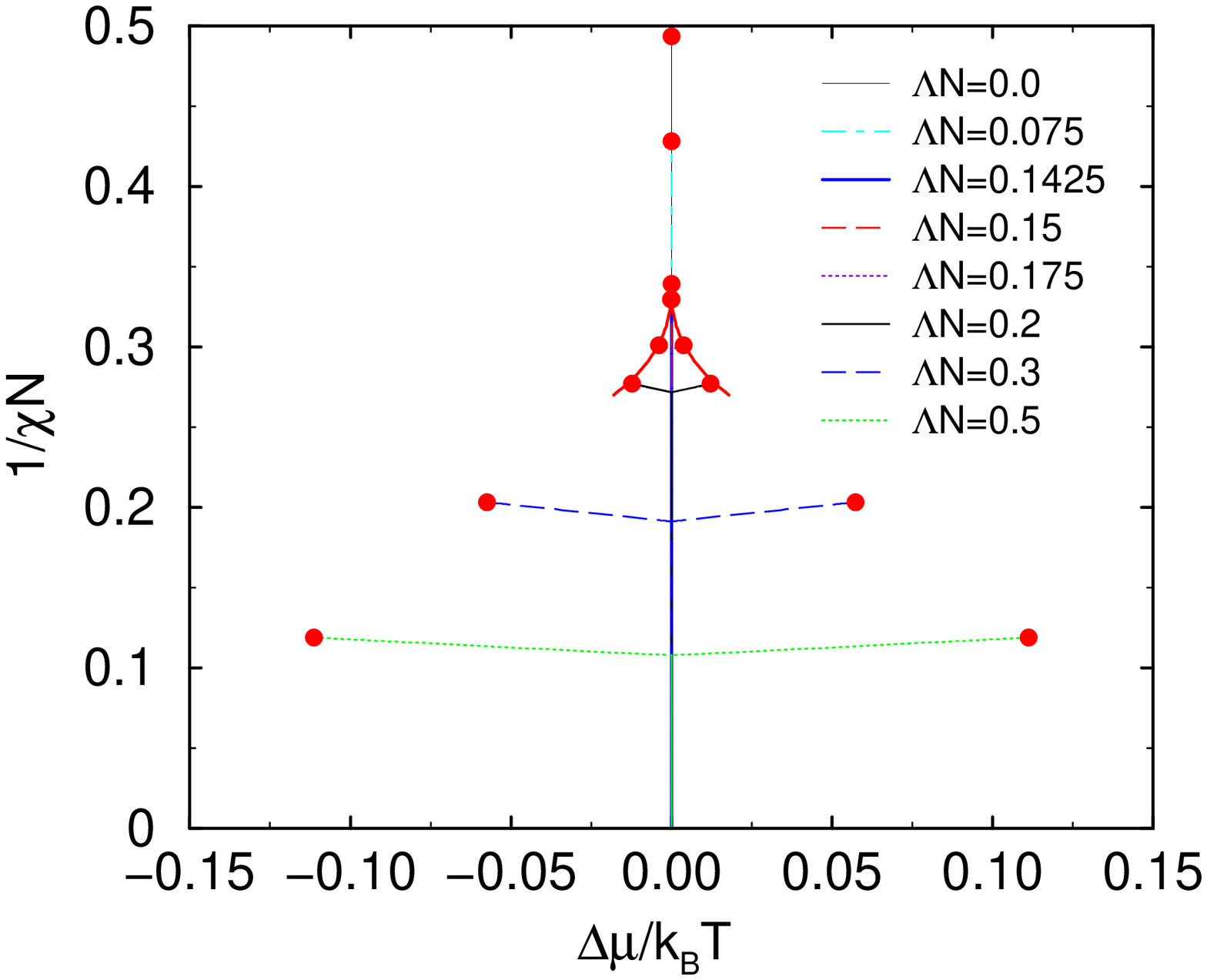}
       }
    \end{minipage}%
    \hfill%
    \begin{minipage}[b]{160mm}%
    \vspace*{1cm}
       \caption{
       \label{fig:D0.9a} 
       Phase diagrams in a thin film ($\Delta_0 = 0.9 R_e$) with antisymmetric surface fields.
       The values of the surface fields $\Lambda N$ are indicated in the key. For
       $\Lambda N\leq 0.1425$ we find a single critical point, while we find two critical points
       for larger surface fields.
       ({\bf a}) displays the phase diagram in the temperature--composition plane, while ({\bf b})
       presents the coexistence curves $\Delta \mu_{\rm coex}(\chi N)$. (For $\Lambda N=0.15$
       the two critical points are indistinguishable on the scale of panel ({\bf b})).
       }
    \end{minipage}%
\end{figure}

The qualitative features of this scenario are confirmed by the SCF calculations. Results for film thickness $\Delta_0 = 0.9 R_e$ are
presented in Fig.\ref{fig:D0.9a}. Panel ({\bf a}) presents the phase diagram for various strength of the surface fields as a function of 
temperature and composition. For weak surface fields $\Lambda N$ the interface localization--delocalization  transition
is of second order and we obtain phase diagrams with one critical point at $\phi_c=1/2$ 
Upon increasing the strength of the surface fields the critical point shifts to lower temperatures. Around $\Lambda N=0.1425$ 
the binodal become flatter and are compatible with an exponent $\beta_{\rm tri}=1/4$. In accordance with the Ginzburg--Landau ansatz this marks 
the tricritical transition. At stronger surface fields we obtain phase diagrams with two critical points, which correspond to the prewetting critical points 
of the first order wetting transition in the semi--infinite system. When we increase $\Lambda N$ further,
the two critical points, which are located symmetrically around $\phi=1/2$, gradually move to lower temperatures, and higher or lower $A$ 
concentration, respectively. Moreover, the temperature distance between the critical points and the triple point increases.

Fig.\ref{fig:D0.9a}({\bf b}) depicts the behavior in terms of temperature and chemical potential difference. For second order interface 
localization--delocalization transitions the coexistence chemical potential is $\Delta \mu_{\rm coex}=0$ by virtue of the symmetry with respect
to exchanging $A \rightleftharpoons B$. The same holds true for first order transitions below the triple point. At the triple point, however, the
coexistence curve bifurcates into two symmetrical branches, which correspond to the prewetting lines of the semi--infinite system.
These lines end at critical points. Upon increasing the strength of the surface fields the two critical points move to lower temperatures and
larger absolute values of $\Delta \mu$.

\begin{figure}[htbp]
    \begin{minipage}[t]{160mm}%
       \mbox{
        \setlength{\epsfxsize}{8cm}
        \epsffile{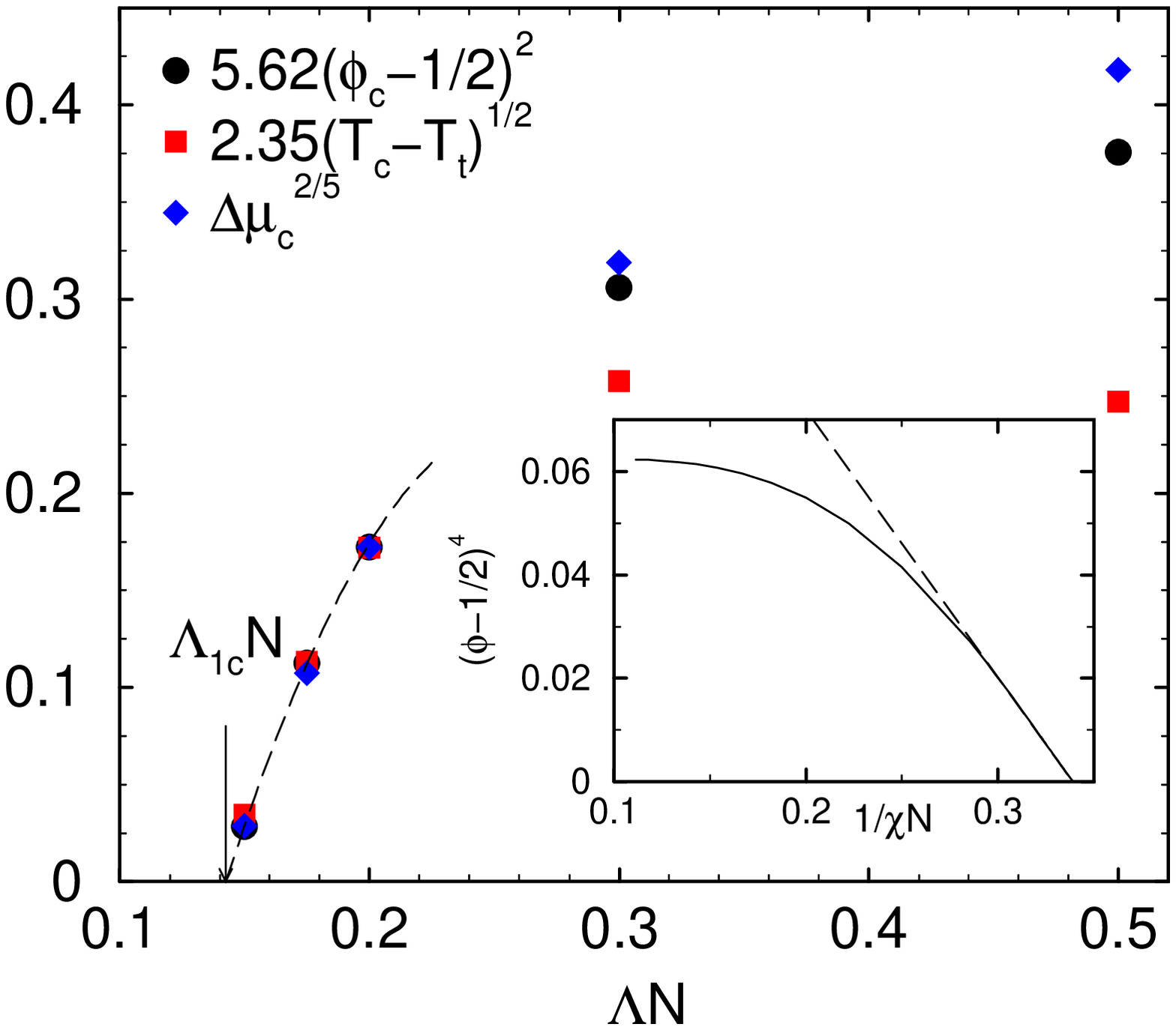}
       }
    \end{minipage}%
    \hfill%
    \begin{minipage}[b]{160mm}%
    \vspace*{1cm}
       \caption{
       \label{fig:crit}}
       \protect{$(\phi_c-1/2)^2$}, \protect{$\sqrt{T_c-T_t}$},  and \protect{$\Delta \mu_c^{2/5}$} 
       as a function of the
       surface field \protect{$\Lambda N$}. In agreement with the Ginzburg--Landau ansatz all 
       quantities
       show the same dependence on the parameter \protect{$r(\Lambda N)$}. The dashed line is only a guide to the 
       eye. The location
       of the critical surface field is indicated. \protect{$\Lambda_{1c}N \approx 0.1425$}. The inset display the
       behavior of the binodals for $\Lambda N = 0.1425 \approx \Lambda_{c}N$. The curve corresponds to the
       SCF calculations, while the dashed line indicates the \protect{$\phi-0.5 \sim \pm |T-T_c|^\beta$} with
       \protect{$\beta=1/4$}.
    \end{minipage}%
\end{figure}

To make closer connection to the Ginzburg--Landau ansatz, we assume that the parameter $r$, which drives the transition
between the two types of phase diagrams, varies as a function of the surface field $\Lambda N$. Then the Ginzburg--Landau
ansatz predicts that the quantities $(\phi_c-1/2)^2 \sim \sqrt{T_c-T_t} \sim \Delta \mu^{2/5} \sim r(\Lambda N)$.
This is tested in Fig.\ref{fig:crit}. Indeed, our SCF calculations confirm that these quantities exhibit a very similar 
dependence on the surface field close to the transition. Moreover, we estimate the critical value of the surface field 
to be $\Lambda N \approx 0.1425$.  The corresponding power laws for the location of the critical points in the vicinity of tricriticality
are also displayed in the Figs.\ref{fig:D0.9a}({\bf a}) and ({\bf b}).
Additionally, the inset of Fig.\ref{fig:crit} shows that the binodals 
are characterized by an exponent $\beta_{\rm tri}=1/4$ at this tricritical value of the surface field. This provides strong 
evidence that the Ginzburg--Landau ansatz captures the salient features of the tricritical transition.

Square gradient calculations\cite{SWIFT} and recent Monte Carlo simulations\cite{B2} of the Ising model indicate that the 
interface localization--delocalization transition
can be second order in thin films ($\Delta_0<\Delta_{\rm tri}$), even if the wetting transition is first order.
Within our Ginzburg--Landau ansatz this finding can be rationalized as follows:
Close to the tricritical point the coefficient $r \sim g_4$ is small and the temperature of the triple point
is given by the condition $0=t \sim g_2 +{\cal O}(r^2)$ or $a_{\rm triple}=4b \exp(-\lambda \Delta_0/2)-9c \exp(-\lambda \Delta_0)
+{\cal O}(r^2)$. At the tricritical film thickness $\Delta_{\rm tri}$ the coefficient $r(t=0)$ changes its sign. Neglecting terms 
of order ${\cal O}(r^2)$ we obtain: $r(t=0) \sim -a+16 b \exp(-\lambda \Delta_0/2)-81c  \exp(-\lambda \Delta_0) \approx
12 b  \exp(-\lambda \Delta_0/2) -72 c   \exp(-\lambda \Delta_0)$. If the semi--infinite system exhibits a first order wetting
transition the coefficients $b$ and $c$ are positive. Hence, for large $\Delta_0$ the coefficient $r$ is positive
and leads to a first order interface localization--delocalization transition. If the
film width $\Delta_0$ becomes comparable to the correlation length $1/\lambda$, however, the second term might drive the
coefficient $r$ negative upon decreasing the film thickness. 

\begin{figure}[htbp]
    \begin{minipage}[t]{160mm}%
       \mbox{
        ({\bf a})
        \setlength{\epsfxsize}{8cm}
        \epsffile{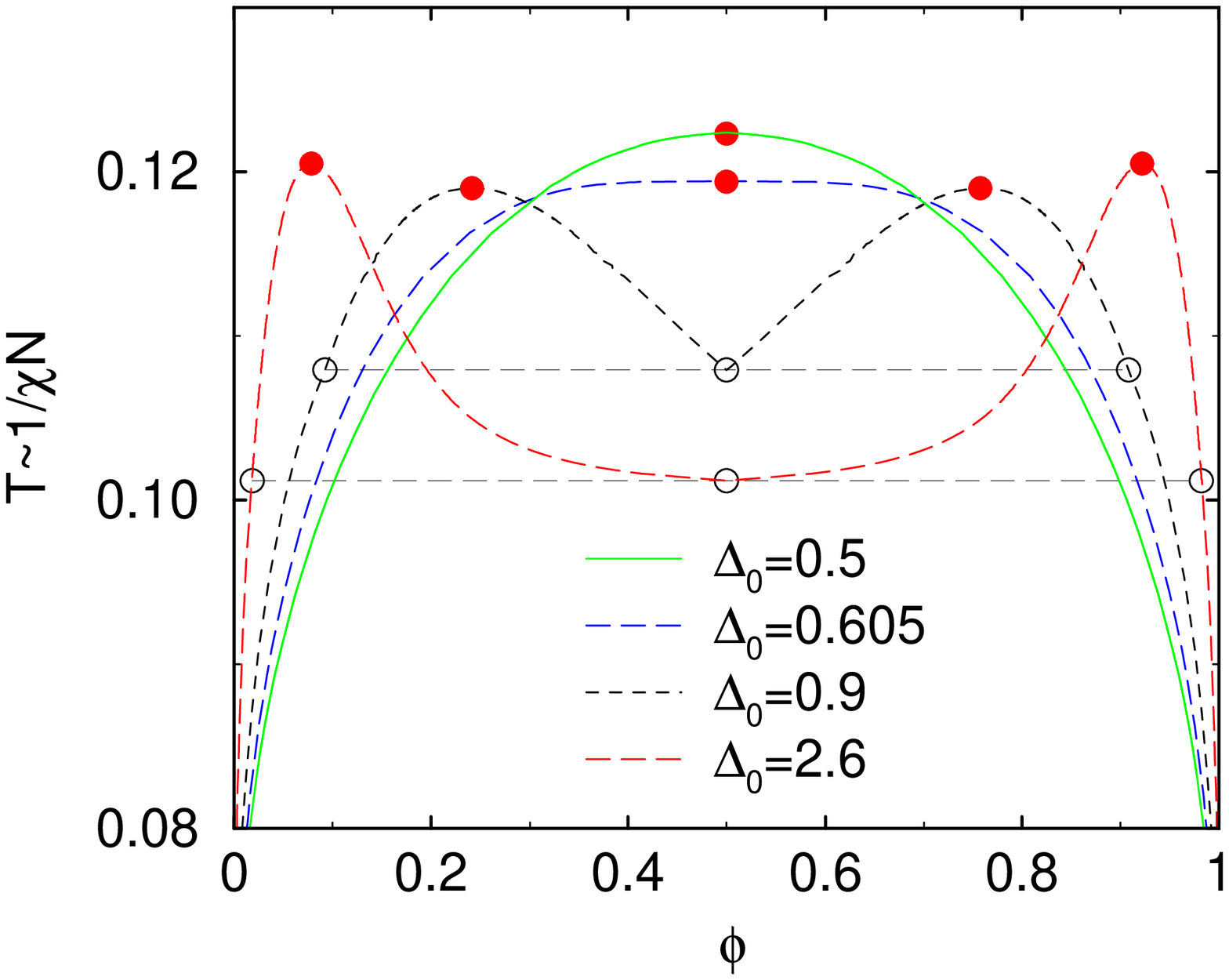}
        ({\bf b})
        \setlength{\epsfxsize}{8.104cm}
        \epsffile{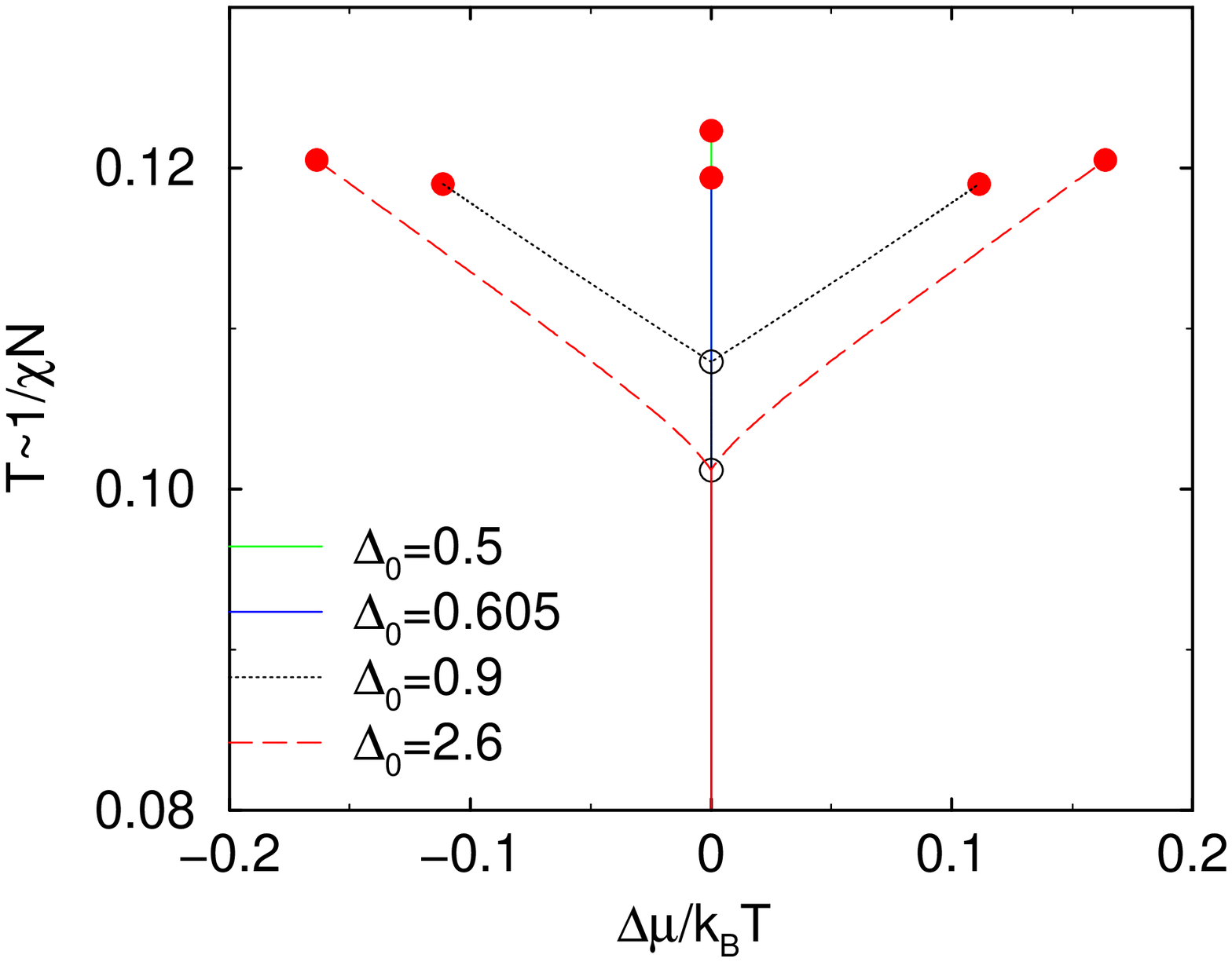}
       }\\
       \mbox{
        ({\bf c})
        \setlength{\epsfxsize}{8cm}
        \epsffile{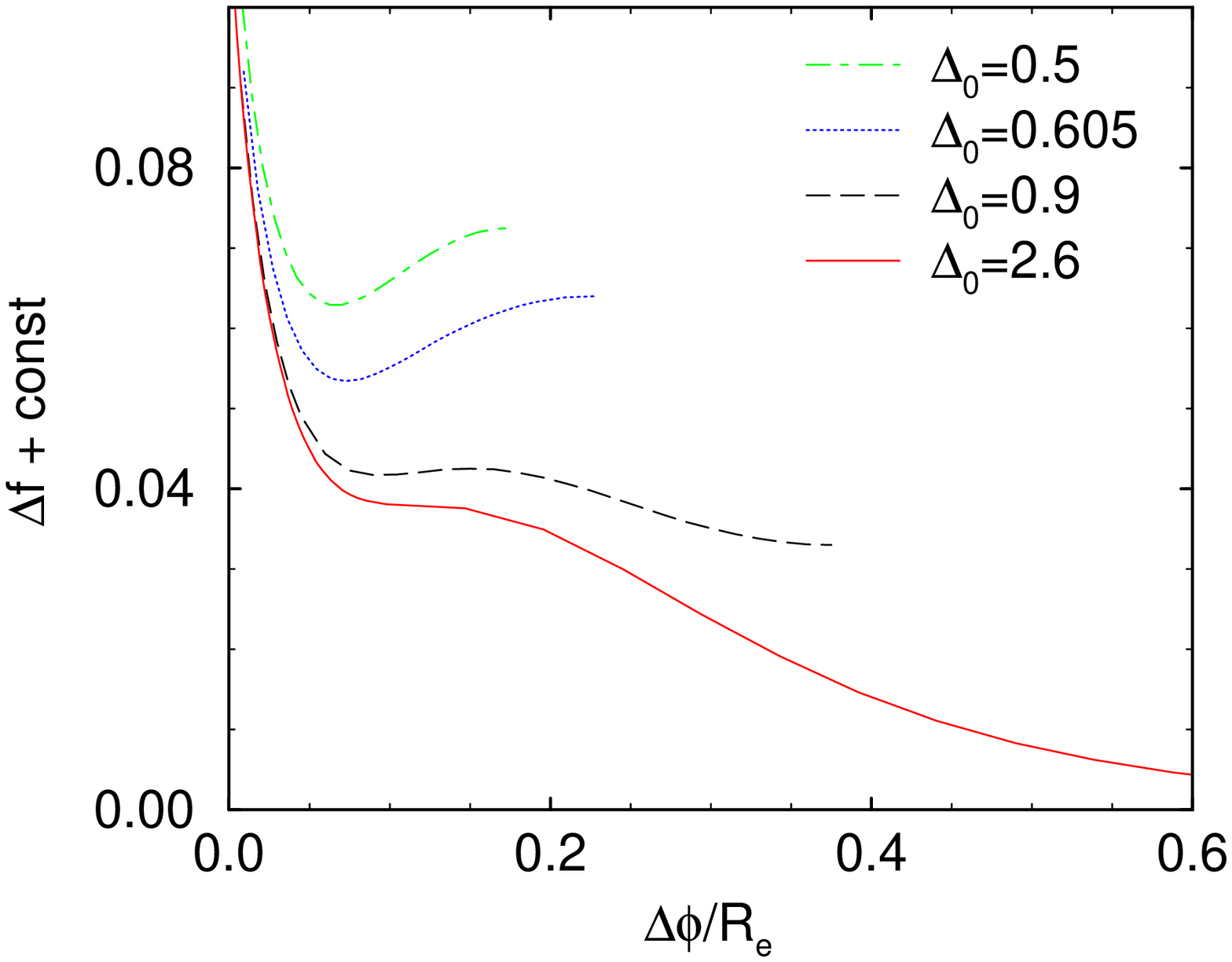}
        ({\bf d})
        \setlength{\epsfxsize}{8.104cm}
        \epsffile{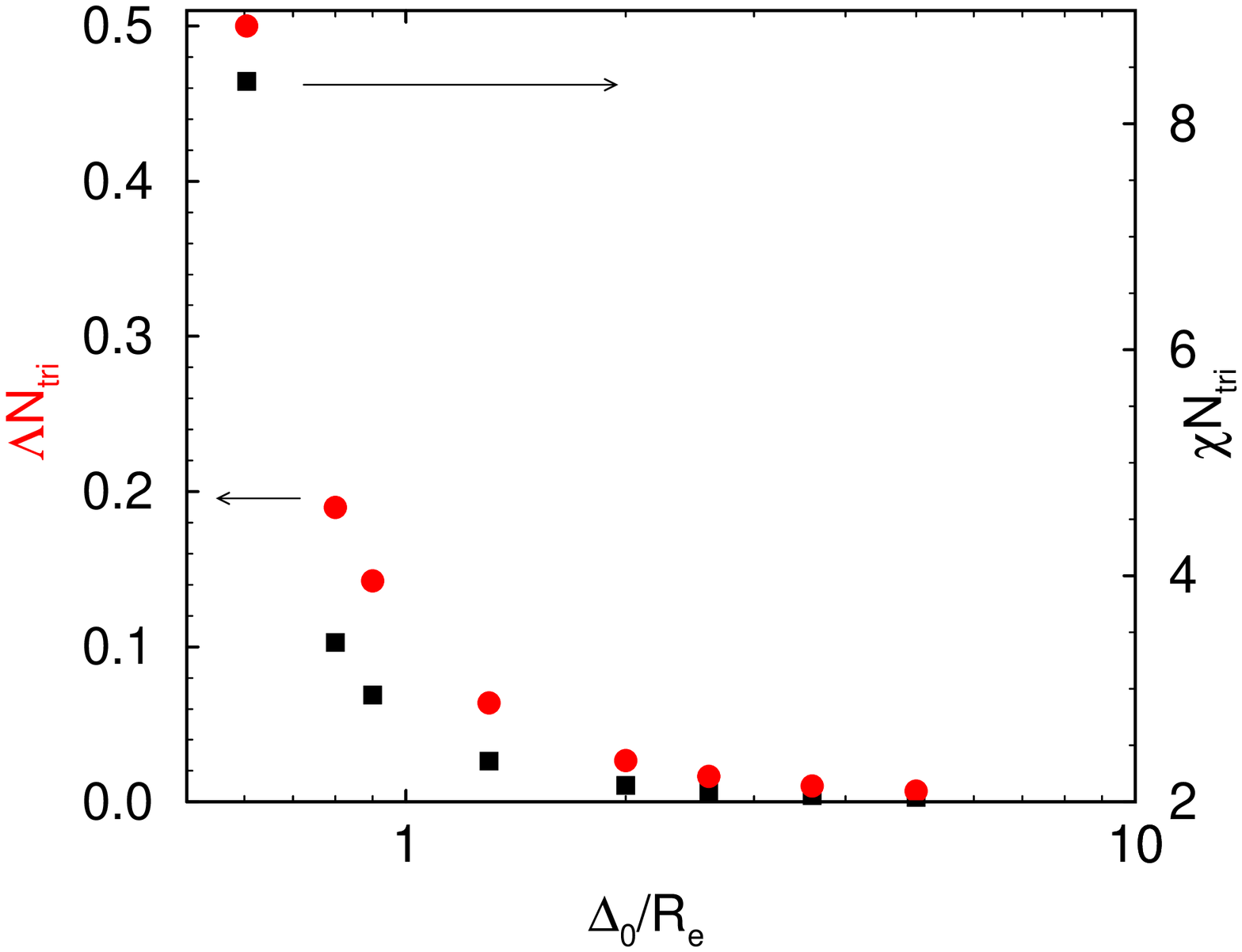}
       }
    \end{minipage}%
    \hfill%
    \begin{minipage}[b]{160mm}%
    \vspace*{1cm}
       \caption{
       \label{fig:thick}}
       ({\bf a}) Phase diagram for \protect$\Lambda N=0.5$ and various film thicknesses \protect$\Delta_0$. For \protect$\Delta_0=2.6 R_e$ and \protect$0.9R_e$
       the interface localization--delocalization transition is first order, \protect$\Delta_0 = 0.605 R_e$ corresponds to a tricritical
       transition, while the transition is second order for \protect$\Delta_0 = 0.5R_e$.
       ({\bf b}) Phase diagram as a function of temperature and chemical potential for the same parameters than in ({\bf a}).
       ({\bf c}) Effective interface potential \protect$g(l) \sim \Delta f(\phi)$ at $l=\Delta \phi$ as a function of the film thickness $\Delta_0$
                 for $\chi N = 9$.
       ({\bf d}) Surface field (circles) and temperature (squares) of the tricritical transition as a function of the film thickness \protect$\Delta_0$.
    \end{minipage}%
\end{figure}

This is further explored in our SCF calculations.
In Fig.\ref{fig:thick}({\bf a}) we present the phase diagrams as a function of the film thickness at $\Lambda N = 0.5$.
For large film thickness $\Delta_0 = 2.6 R_e$, we find a first order interface localization--delocalization transition.
Upon decreasing the film thickness the two critical points move to lower temperatures and closer to the symmetry axis.
$\Delta_0 \approx 0.605$ corresponds to the tricritical transition: There is only a single critical point but the binodals are
describable by the exponent $\beta_{\rm tri}=1/4$. Upon further decreasing $\Delta_0$ the critical temperature increases and the 
binodals assume parabolical shape. The corresponding coexistence curves are shown in Fig.\ref{fig:thick}({\bf b}). 

The mechanism is most clearly visible in the effective $AB$ interface potential $g(l) \sim \Delta f$, which is presented in Fig.\ref{fig:thick}({\bf c}) for $\chi N = 9$
as a function of the distance $l = \Delta \phi$ between the surface and the $AB$ interface. For the largest film thickness $\Delta_0=2.6 R_e$
$g(l)$ is to a good approximation the effective potential of the nearest surface $g_{\rm wall}$. $\chi N = 9$ corresponds to a temperature
above the wetting (triple) transition but below the (prewetting) critical temperature of the thick film. Each surface repels the $AB$ interface 
and there is a shoulder in the effective interaction around $l_{\rm prewet} \approx 0.22 R_e$.
Qualitatively, the SCF calculations confirm that $g(l)$ is the linear superposition of the interactions with each surface. When we decrease the
film thickness larger values of $l$ become unfavorable, because the more the interface moves away from one surface the more it experiences 
the repulsion from the opposite surface. This results in a minimum of $g(l)$ close to each surface. The shape of $g(l)$ corresponds to a 
temperature below the triple point, i.e., the interface is localized at one or the other surface. This shows that the triple temperature
increases when we decrease the film thickness. The first order character of the transition is associated with the shoulder of $g_{\rm wall}$ around 
$l_{\rm prewet}$. Obviously, this feature of $g(l)$ disappears when the film thickness $\Delta$ is of the order $2l_{\rm pw}$ and we find second order
transitions for smaller film thicknesses.

This thickness dependence implies that the value $\Lambda_{\rm tri} N = 0.57$ obtained from the phase diagram of rather thin films $\Delta_0 = 0.9 R_e$
is not a reliable approximation of the strength of the surface fields at which a tricritical wetting transition in the semi--infinite system occurs.
We have attempted to locate the strength of the surface fields at which the binodals are describable by the exponent $\beta_{\rm tri}=1/4$ as a function of the
film thickness $\Delta$.
The results of this procedure are collected in Fig.\ref{fig:thick}({\bf d}). As we increase $\Delta$ the surface fields at which the tricritical 
transition occurs decreases and the transition temperature approaches the critical point. For film thickness much larger than the range of interaction
$1/\lambda$ between the surface and the $AB$ interface we expect a thin film to behave similar as a semi--infinite system.
Close to the critical point, however,  the range of interaction $1/\lambda = \xi$ between the surface and the $AB$ interface increases. Therefore
we anticipate very pronounced finite size effects even for film thicknesses which exceed the end--to--end distance $R_e$ by far. These difficulties
prevent us from reliably estimating the tricritical wetting transition of the semi--infinite system or comparing our calculations to the prediction (\ref{eqn:2nd})
of the square--gradient approximation.\cite{SG1,SG2} From the behavior at film thickness $\Delta_0=5R_e$ we conclude that critical wetting transitions in the semi--infinite systems
occur only for $\Lambda N < 0.01 $ and $\chi N < 2.04$ in our model. Qualitatively this is in agreement with SCF calculations of Carmesin and Noolandi\cite{CN}
and Monte Carlo simulations\cite{WET} which find only first order wetting transitions, except for the ultimate vicinity of the critical point which has
not been investigated.

\subsection{Interfacial profiles}
The effective interface potential also determines the composition profiles across an interface between the laterally coexisting phases.
At low temperatures the coexisting phases are almost completely segregated, i.e., the thickness of the enrichment layers of the minority components are
small. In this case the interface between the coexisting phases is planar and makes an angle $\Theta$ with the surface. The contact angle $\Theta$
is given by the Young equation:
\begin{equation}
\cos \Theta = \frac{\sigma_{WA}-\sigma_{WB}}{\sigma_{AB}} \approx \frac{\Lambda N}{\sqrt{\chi N/24}} \qquad \mbox{(SSL)}
\end{equation}
The width $w$ of the interface between the coexisting phases is given by $w = \Delta_i {\rm ctg}\Theta$, where $\Delta_i/2$ denotes the 
distance of the $AB$ interface from the center of the film in each phase. At a second order interface localization--delocalization transition 
the contact angle decreases linearly with the distance from the transition temperature $t$ and the composition difference between the coexisting 
phases vanishes like $\Delta_i \sim |t|^{1/2}$. Hence, the width of the interface between the coexisting phases diverges like $w \sim |t|^{-\nu}$
where $\nu=1/2$ is the mean field exponent for the correlation length upon approaching the critical point.

Close to a first order interface localization--delocalization transition the effective interface potential exhibits more structure and 
this will modify the shape of the interface. Within the mean field approximation the shape of the interface will minimize the effective
interface free energy. Approximating the $AB$ interface profile as a sharp kink at the position $l$ we obtain for the free energy the
effective interface Hamiltonian:\cite{IH}
\begin{equation}
\frac{F[l]}{k_BT} = \int {\rm d}y{\rm d}z\; \left\{ \sigma_{AB} \left(\sqrt{1+ \left(\frac{{\rm d}l}{{\rm d}y}\right)^2}-1 \right) + g(l) \right\}
 \approx L \int {\rm d}y\; \left\{ \frac{\sigma_{AB}}{2}  \left(\frac{{\rm d}l}{{\rm d}y}\right)^2 + g(l) \right\}
 \label{eqn:H}
\end{equation}
where we have assumed that the position $l$ of the $AB$ interface depends only on one lateral coordinate $y$. The last approximation is
valid if the angle between the interface and the surface is small. This is justified in the vicinity of the wetting transition, but the
approximation breaks down at low temperatures, where the interface runs almost perpendicular to the surfaces. This effective interface Hamiltonian 
yields the Euler--Lagrange equation:
\begin{equation}
\frac{{\rm d}^2l}{{\rm d}y^2} = - \frac{{\rm d}}{{\rm d}l} \left( -\frac{g(l)}{\sigma_{AB}}\right)
\label{eqn:EL}
\end{equation}
which can be interpreted as the trajectory $l$ of a particle in the potential $-g/\sigma_{AB}$. To obtain 
a qualitative insight we extract the effective interface potential from the one--dimensional SCF calculations
$g(l)/\sigma_{AB}=k_BT \sqrt{\bar N} \Delta f(\phi_A)/\sigma_{AB}R^3$ at $l=\Delta \phi_A$.

\begin{figure}[htbp]
    \begin{minipage}[t]{160mm}%
       \mbox{
        \setlength{\epsfxsize}{7cm}
        \epsffile{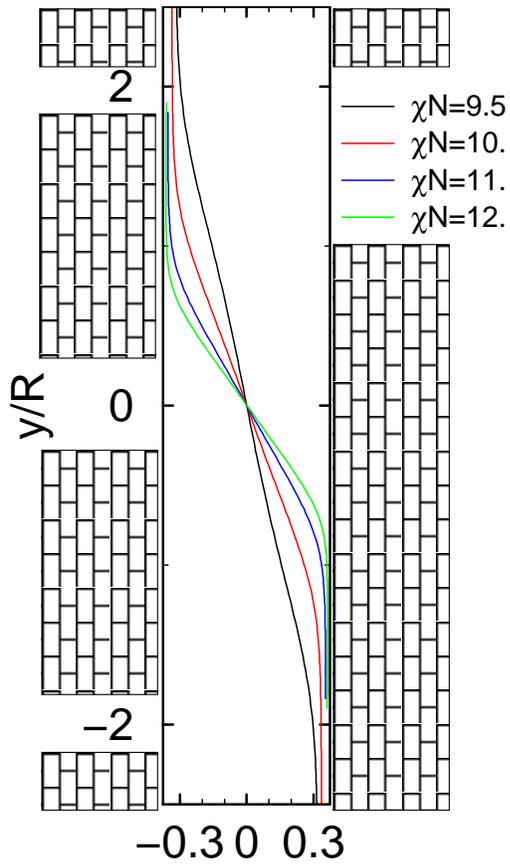}
       }
    \end{minipage}%
    \hfill%
    \begin{minipage}[b]{160mm}%
    \vspace*{1cm}
       \caption{
       \label{fig:prof}
       Shape of the interface between the coexisting $A$--rich and $B$--rich phases below the triple point.
       The profiles have been obtained from the Eq.(\protect\ref{eqn:EL}) using input from the one--dimensional
       SCF calculations.
       }
    \end{minipage}%
\end{figure}

Typical shapes of the interface between the coexisting phases for a film thickness $\Delta_0=0.9 R_e$ are presented in 
Fig.\ref{fig:prof}. Far below the triple temperature the interface is planar and makes an angle $\Theta$ with the
surfaces. Slightly below the triple temperature, however, the interfacial profile becomes $s$--shaped, i.e.,
the angle between the interface and the surface is larger in the vicinity of the surface than at the center of the film.
Note that the lateral interfacial width can exceed the film thickness by far in the vicinity of the triple temperature.
The flatter portion in the center of the film is a consequence of the metastability of the third phase with composition 
$\phi=1/2$ or the additional local minima in $g(l)$, respectively. Upon approaching the triple temperature the interfacial width
becomes larger and  the central portion of the profiles becomes flatter and more extended. This central portion might be conceived 
as a microscopic layer of the metastable delocalized phase ($\phi=1/2$), which completely wets the interface between the $A$-rich and 
$B$--rich phase at the triple point.

In the semi--infinite system the interfacial tension varies smoothly upon rising the temperature through
the wetting transition temperature. The excess free energy of the interface approaching the surface -- the line tension $\tau$ --
varies rapidly close the wetting transition. Employing an effective interface Hamiltonian of the form (\ref{eqn:H}), Indekeu\cite{IND} 
has obtained a simple expression for the line tension $\tau$.
\begin{equation}
\tau = \sqrt{2 \sigma_{AB}} \int_{l_1}^\infty {\rm d}l\; \left\{ \sqrt{\delta g_{\rm wall}(l)} - \sqrt{\delta g_{\rm wall}(\infty)}\right\}
\label{eqn:ind}
\end{equation}
where $l_1$ is the position of the minimum of $g_{\rm wall}(l)$ close to the surface and $\delta g_{\rm wall}=g_{\rm wall}(l)-g_{\rm wall}(l_1)$.
This formula has been applied to analyze recent experiments.\cite{WANG} The behavior of the line tension close to the
wetting transition depends on the order of the transition and the range of the monomer--wall interaction. For short range forces
the line tension $\tau$ reaches a finite positive value at the wetting transition temperature, while it is negative far below
the wetting temperature.

\begin{figure}[htbp]
\mbox{
    \hspace*{-1cm}
    \begin{minipage}[c]{40mm}%
        \setlength{\epsfxsize}{4.3235cm}
        \epsffile{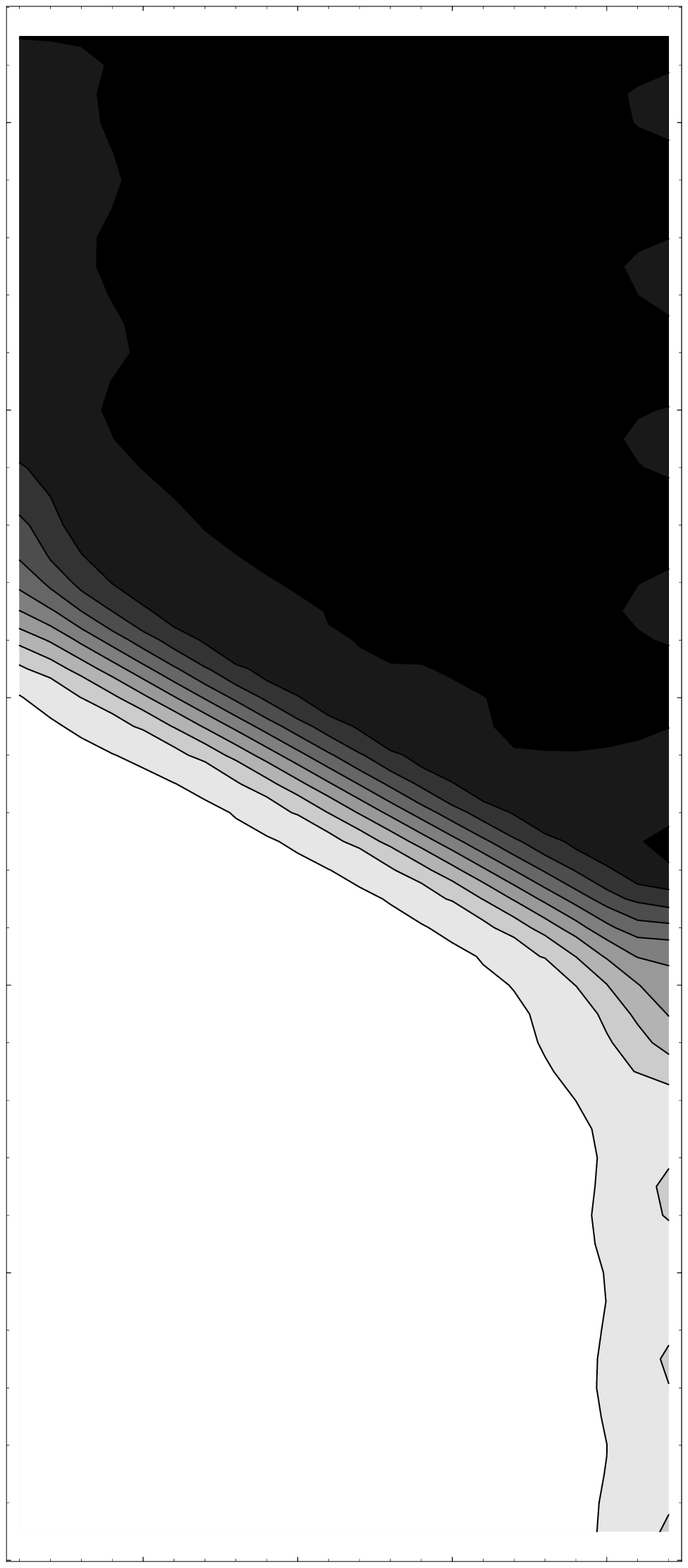}
    \end{minipage}
    \hspace*{-1cm}
    \begin{minipage}[c]{40mm}%
        \setlength{\epsfxsize}{4.3235cm}
        \epsffile{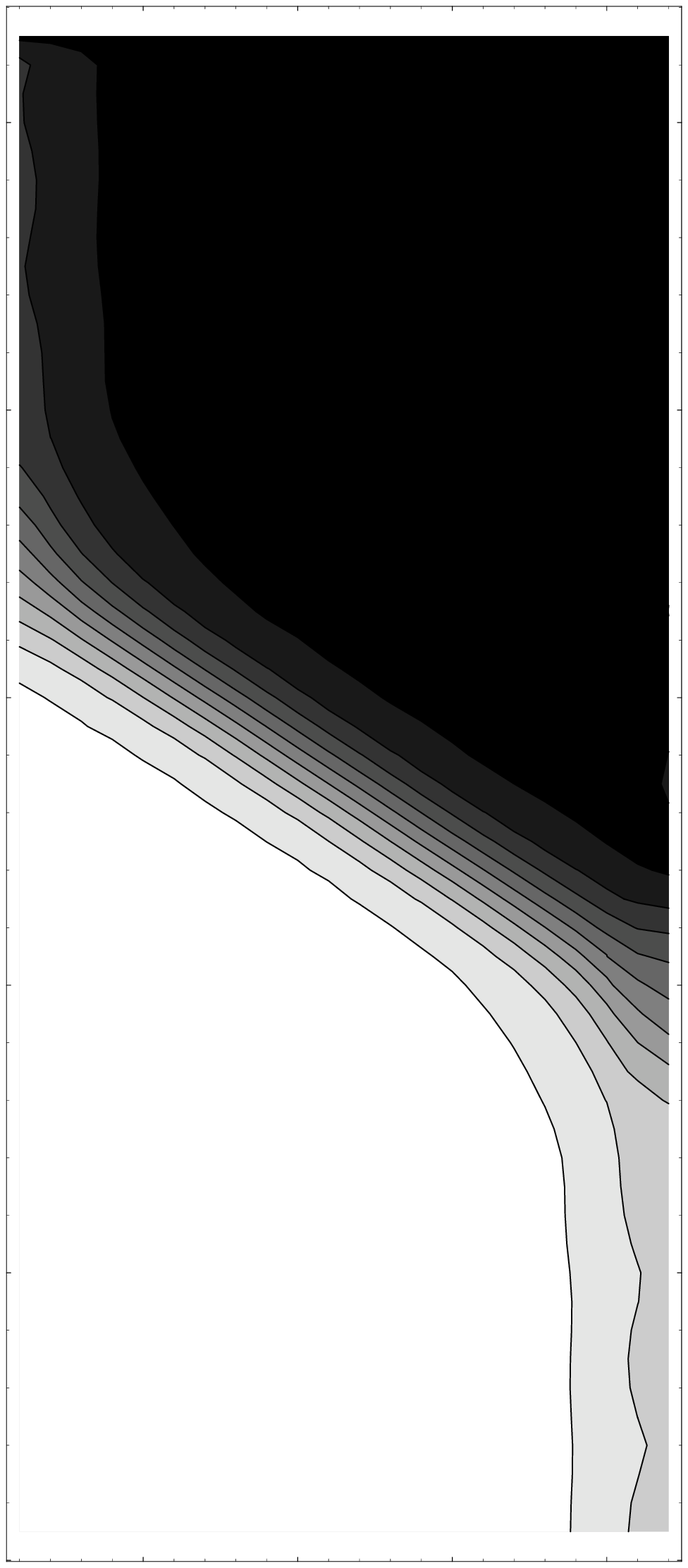}
    \end{minipage}
    \hspace*{-1.3cm}
    \begin{minipage}[c]{40mm}%
        \setlength{\epsfxsize}{4.94cm}
        \epsffile{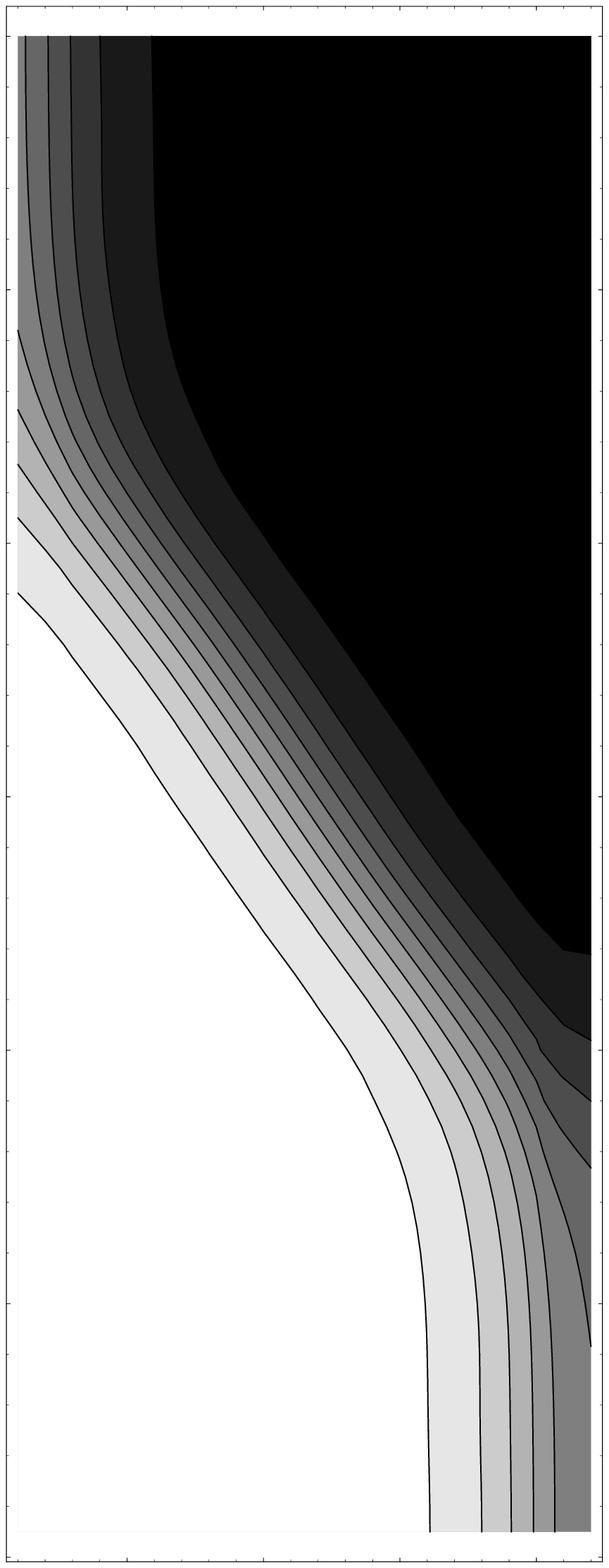}
    \end{minipage}
    \hspace*{-2.7cm}
    \begin{minipage}[c]{40mm}%
        \setlength{\epsfxsize}{7cm}
        \epsffile{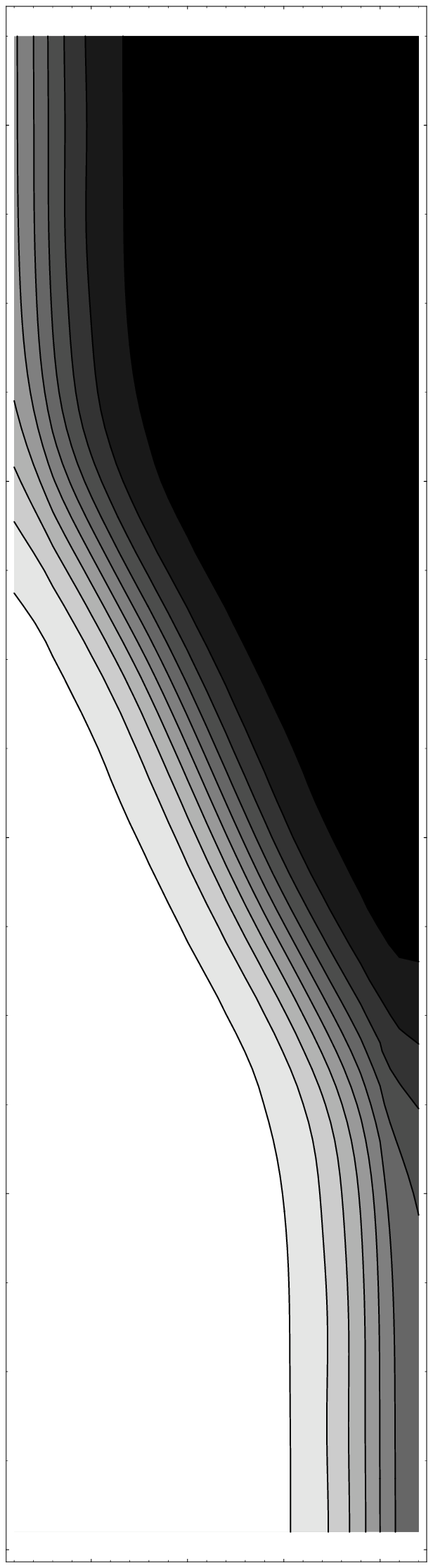}
    \end{minipage}
    \hspace*{-3.6cm}
    \begin{minipage}[c]{40mm}%
        \setlength{\epsfxsize}{7cm}
        \epsffile{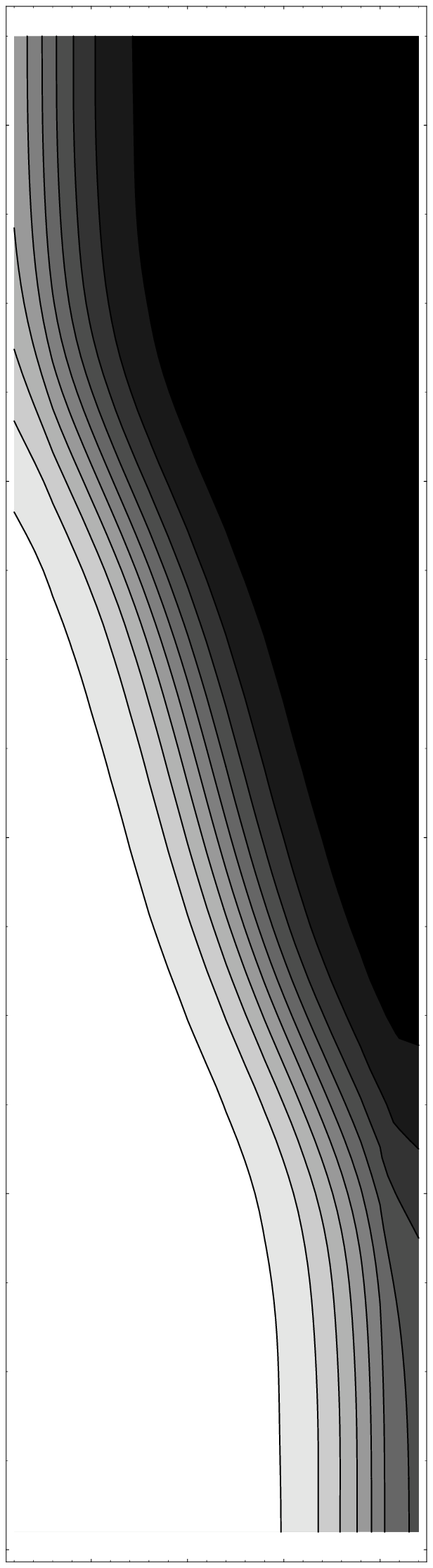}
    \end{minipage}
}
    \begin{minipage}[b]{160mm}%
    \vspace*{1cm}
       \caption{
       \label{fig:profs}
       Composition profiles of the interface between the $A$--rich and $B$--rich phase at
       $\chi N = 13, 12, 10, 9.63$, and $9.44$ and film thickness $\Delta_0=0.9 R_e$ and lateral
       extension $L/2=5.25, 6$ and $8.5 R_e$. For clarity the aspect ratio of the figure has 
       been decreased.
       }
    \end{minipage}%
\end{figure}

The effective interface Hamiltonian captures only the qualitative behavior. Monte Carlo simulations and SCF calculations have
shown that the properties of the $AB$ interface depend on the distance $l$ from the surface. This gives rise to a position 
dependence of the tension\cite{WET} and width of the $AB$ interface.\cite{ANDREAS}
A more detailed description of the interface is provided by the two--dimension
composition profiles in Fig.\ref{fig:profs}. Due to the choice of basis functions the profiles are periodic in $y$ direction 
and only half the system is shown. In qualitative agreement with the considerations above, the interface between the 
$A$--rich and $B$--rich phases runs straight across the film at low temperatures ($\chi N = 13$ and $12$). The contact angle 
at $\chi N = 12$ is about $30^0$.

Upon increasing the temperature the contact angle between the surface and the interface decreases and the interface becomes $s$--shaped
in the vicinity of the triple point. The SCF calculations also reveal that the interface becomes broader when we increase the temperature.
Moreover, the width of the interface is broader in the vicinity of the surfaces than in the middle of the film. This effect is due to the reduction
of the monomer density
in the vicinity of the surface, which imparts a reduced effective incompatibility (``missing neighbor effect'') on the surface region. A similar
effects has been observed in confined systems containing copolymers\cite{SCF4,SCF5} This effect gives rise to a negative contribution to the 
line tension when the interface approaches the surface.

\begin{figure}[htbp]
    \begin{minipage}[t]{160mm}%
       \mbox{
        \setlength{\epsfxsize}{8cm}
        \epsffile{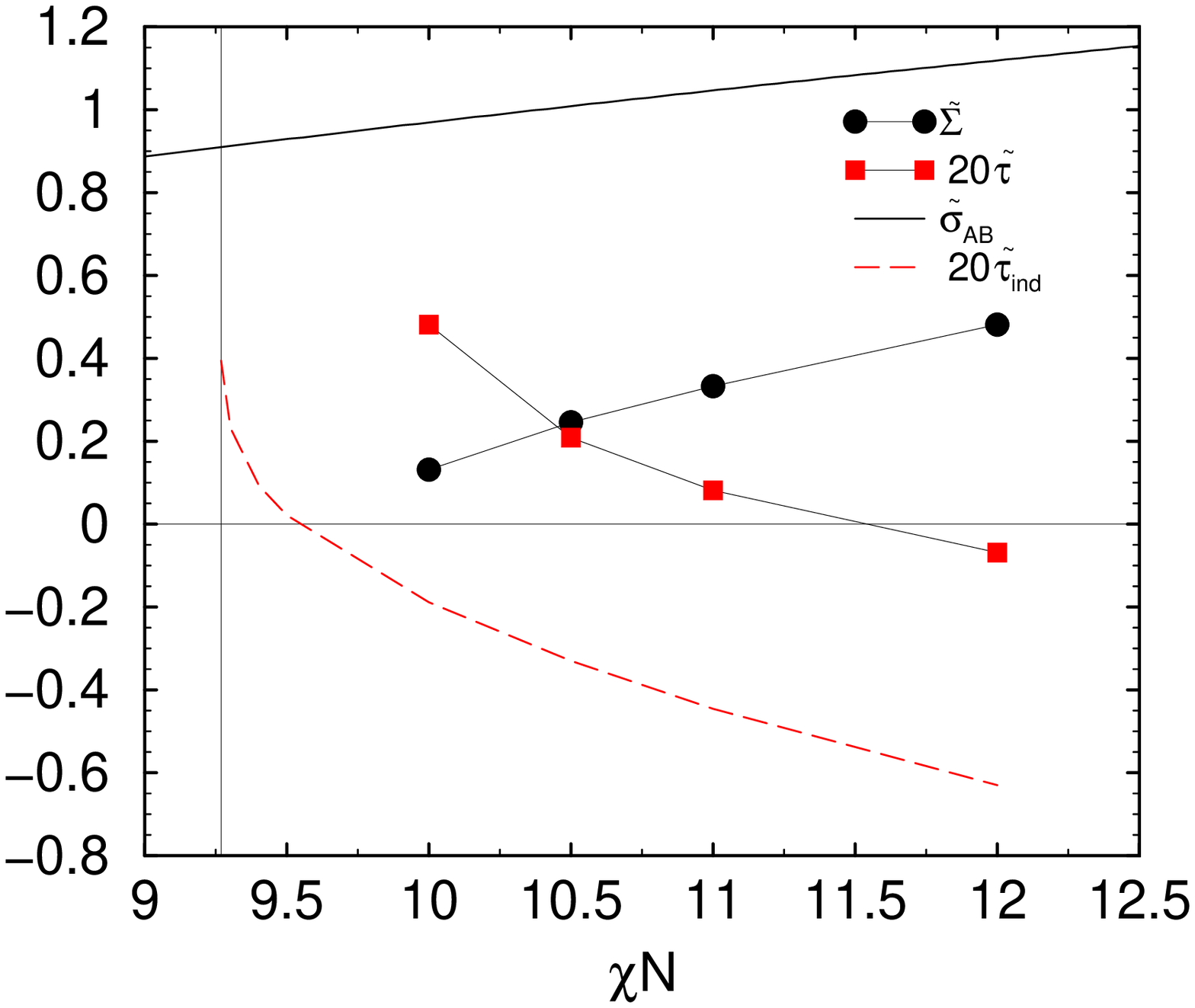}
       }
    \end{minipage}%
    \hfill%
    \begin{minipage}[b]{160mm}%
    \vspace*{1cm}
       \caption{
       \label{fig:line}
       Behavior of the line tension $\tilde \tau$ upon approaching the triple point. Squares denote the line tension as extracted
       from two-dimensional SCF calculations, the dashed line corresponds to Indekeu's approximation applied to a thin film.
       The coefficient $\time \Sigma$ and the $AB$ interface tension $\sigma_{AB}$ are displyed as circle and full line,
       respectively. The vertical lines marks the triple temperature.
       }
    \end{minipage}%
\end{figure}

We decompose the free energy of systems containing two interfaces between an $A$--rich and a $B$--rich phase into bulk, surface and line contributions:
\begin{equation}
F-F_{\rm bulk} = 2(L\Delta \Sigma+2 L \tau) \qquad \mbox{or} \qquad f-f_{\rm bulk}(\Delta) = \frac{2R_e}{L} \tilde \Sigma + \frac{4R_e^2}{L\Delta} \tilde{\tau}
\qquad \mbox{with} \qquad \tilde{\Sigma} = \frac{\Sigma R_e^2}{\sqrt{\bar N}k_BT} \qquad  \tilde{\tau} = \frac{\tau R_e}{\sqrt{\bar N}k_BT} 
\end{equation}
Note that the surface fields and the entropy loss of the chains at the surfaces give rise to a thickness dependence of the bulk
free energy density $f_{\rm bulk}$. Varying both $L$ and $\Delta$ we have estimated the coefficients in our SCF calculations and the results are displayed 
in Fig.\ref{fig:line}. Qualitatively similar to the behavior at a first order wetting transition the line tension changes its sign from negative to positive 
upon approaching the triple point temperature from below. The coefficient $\tilde{\Sigma}$ decreases as we approach the triple point. To a first 
approximation one would expect a behavior of the form $\Sigma \approx \sigma_{AB} \Theta$.

The data are also compared to Indekeu's formula (\ref{eqn:ind}). In order to apply the formula to thin films, we extend the integration only to the
middle of the film and we shift the constant $\delta g_{\rm wall}(\infty)$ accordingly to $\delta g_{\rm wall}(\Delta/2)$
In the dimensionless units of the SCF calculations we obtain:
\begin{equation}
\tilde{\tau} \approx \sqrt{ 2 \left(\frac{\Delta}{R_e}\right)^3 \frac{\sigma_{AB}R_e^2}{\sqrt{\bar N} k_BT} } \int_{\phi_1}^{1/2} {\rm d} \phi \;
\left\{ \sqrt{f(\phi)-f(\phi_1)} - \sqrt{f(1/2)-f(\phi_1)} \right\}
\end{equation}
This approximation for a thin film gives a reasonable estimate for the temperature dependence and the order of magnitude of the line tension; however, 
the value of the lines tension in a thin film is systematically underestimated. When we apply the above approximation to thicker films the value of
the line tension increases. If we use the film thickness $\Delta_0=2.6 R_e$ instead of $\Delta_0=0.9 R_e$ the approximation yields $\tilde \tau = -0.016$
instead of $\tilde \tau = -0.032$. Unfortunately, we are unable to extend our SCF calculations to larger film thicknesses. 

In addition to the
finite film thickness there are other effects which might upset the comparison between the SCF calculations and the effective Hamiltonian description:
It is unclear how accurate the identification of the effective interface potential via $g(l)= \rho k_BT \Delta f(\phi=l/\Delta)/N$ is. 
This identification of the interface position via the absorbed amount is a good approximation for large distances between the surface 
and the $AB$ interface. In this case, the composition profile across the $AB$ interface is well describable by the interfacial profile between 
the coexisting bulk phases. If the $AB$ interface is close to the surface, however, the profile becomes strongly distorted (cf.\ Fig.\ref{fig:gfree}) 
and the definition of the interface position $l$ is somewhat ambiguous, but this is exactly the region which gives the dominant contribution to the line 
tension. 
Moreover there are non--local contributions to the free energy, e.g., due to the conformational entropy. The polymers change their conformation
as to fill the wedge--shaped volume between the surface and the $AB$ interface. This differ from the behavior close to a surface or an $AB$ 
interface and gives rise to a contribution to the line tension, which is only partially described by the effective interface Hamiltonian.

\section{ Summary and discussion. }
We have calculated the phase diagram of a symmetric polymer mixture
confined to a thin film in mean field approximation. The left surface attracts the $A$ component
with the same strength than the right surface the $B$ component of the mixture. The calculations 
reveal a rich interplay between the phase behavior in confined geometry and the wetting behavior 
of the semi--infinite system. If the wetting transition of the semi--infinite system is second
order so is the interface localization--delocalization transition in a thin film.\cite{PE}

At stronger surface fields the wetting transition in the semi--infinite system is first order
and this gives rise to a first order interface localization--delocalization transition in a thick
film. The phase diagram in a thin film exhibits two critical points symmetric around $\phi=1/2$.
These correspond to the prewetting critical points of the semi--infinite system. At lower temperatures
we encounter a triple point at which an $A$--rich phase, a phase where the $AB$ interface is located
in the center of the film, and a $B$--rich phase coexist. This triple temperature converges from above 
to the wetting temperature as we increase the film thickness. Below the triple temperature there is
a single coexistence region between an $A$--rich and a $B$--rich phase. The interplay between the prewetting
behavior and the phase diagram in a thin film has been considered for symmetric surface fields (capillary condensation),\cite{WET,NAKANISHI,TRIPLE1}
but -- to the best of our knowledge -- phase diagrams with two critical points far below the bulk critical temperatures 
in films with antisymmetric surface fields have neither been discussed analytically\cite{PRE1,PRE2} nor observed in experiments or simulations.
As we shall discuss below, we do not expect corrections to the mean field calculations to alter our conclusions qualitatively
and we hope our predictions to be confirmed by experiments or simulations.

Qualitatively, the interplay between the prewetting behavior and the phase diagram in a film with antisymmetric boundaries is not specific to polymer blends
but is rather characteristic of all binary mixtures. Symmetric polymer mixtures might, however, be especially suitable model systems for exploring these effects
experimentally and we hope our detailed calculations to provide some guidance.

The existence of the triple point also influences the shape of the interface between the laterally segregated, coexisting phases.
At low temperatures the interfaces runs straight across the film and the angle between the interface and the surface is given by the
macroscopic contact angle. Upon approaching the triple point from below, however, the profiles becomes s--shaped with a flatter portion
at the center of the film. This signals the metastability of the third delocalized phase with composition $\phi=1/2$. At the triple
point the delocalized phase completely wets the interface between the $A$--rich and $B$--rich phase. Upon approaching the triple temperature
from below the line tension changes sign from negative to positive. The properties of the interface between the coexisting phases are in qualitative
agreement with the results on an effective Hamiltonian description.

In thin films the interface localization--delocalization transition might be of second order
even though the wetting transition is of first order. This has been predicted in the framework of 
a square--gradient approach by Swift {\em et al.},\cite{SWIFT} and is in accord with simulations of the Ising 
model.\cite{B2} A similar behavior is found in our self-consistent field calculations for polymer blends.
For our model second order wetting transitions are restricted to the ultimate vicinity of the critical 
point of the bulk while second order interface localization--delocalization transitions can be observed 
far below the bulk critical temperature for film thickness comparable to the end--to--end distance $R_e$. 
This leads us to anticipate very strong finite film thickness effects close to a second order wetting transition
even for film thicknesses which exceed $R_e$ by far.

Of course, our self-consistent field calculations neglect fluctuations.
In the vicinity of the critical point we rather expect 2D Ising critical 
behavior with much flatter binodals than the parabolic binodals of the
mean field universality class. The Ginzburg criterion ensures, however,
that these composition fluctuations are only important in the ultimate
vicinity of the critical point $|1 - \chi_c N/\chi N| \sim 1/\bar{N}$,
where the reduced chain length $\bar{N}=(\rho R_e^3/N)^2$  measures the degree
of interdigitation.  As we increase  $\bar{N}$ the relative temperature distance 
from the critical point for which these composition fluctuations are important 
decreases.  For small and intermediate chain lengths an interesting interplay 
between mean field, 3D Ising and 2D Ising critical behavior is anticipated.
In the limit of large interdigitation, however, we expect composition fluctuations 
to be only of minor importance for most part of the phase diagram.

Moreover, the interface profiles in the self-consistent field calculations
are ideally flat, i.e., there are no capillary waves of the interfaces. The
importance of these fluctuations is not restricted to the vicinity of critical 
points. On the one hand, capillary waves lead to a broadening of profiles across 
the interface. The ``internal'' $AB$ interfaces run parallel to the surfaces and
the effective interaction between the $AB$ interface and the surface imparts
a long wavelength cut--off  $\xi_\|$ to the spectrum of capillary waves. 
Hence, the interfacial width does not grow unbound as we increase the lateral 
system size, but still the self-consistent field calculations might severely 
underestimate the width of the ``internal'' $AB$ interfaces.\cite{KLEIN,SF,ANDREAS} Within a convolution
approximation the apparent width $w_{\rm cap}$ of the $AB$ interface, which is
observed in experiments or simulations, is related to the intrinsic width $w_{\rm scf}$ 
in the SCF calculations via:
\begin{equation}
\left( \frac{w_{\rm cap}}{w_{\rm scf}} \right)^2 
\approx  1 +\frac{k_BT}{4 \sigma_{AB} w_{\rm scf}^2} \ln \left( \frac{\xi_\|}{B}\right) 
\approx 1+ \frac{3\sqrt{6}}{2} \frac{\sqrt{\chi N}}{\sqrt{\bar N}} \ln \left( \frac{\xi_\|}{B}\right)\qquad \mbox{(SSL)}
\end{equation}
where we have used the temperature dependence of the interfacial tension and width for 
strong segregation to obtain the last expression.\cite{SCF1,SEM} $B$ is a short length cut--off for the 
capillary wave spectrum. Analytical calculations and recent Monte Carlo simulations show 
that $B$ tends to $\pi w_{\rm scf}$\cite{SEM2} or $3.8  w_{\rm scf}$,\cite{ANDREAS2} respectively,  in the strong
segregation limit. Even though the second term in the above equation is only of the order 
$1/\sqrt{\bar{N}}$ the increase of the apparent interfacial width due to capillary
waves typically is of the order of the intrinsic width for experimentally relevant chain
lengths.\cite{RUSSELL,KLEIN,SF}

On the other hand, capillary waves renormalize the effective interaction between the surface 
and the $AB$ interface.  For instance, the effective interaction range $1/\lambda$ is increased 
to $(1+\omega/2)/\lambda$, where the capillary parameter\cite{OMEGA1,OMEGA2}
\begin{equation}
\omega=\frac{k_BT \lambda^2}{ 4\pi\sigma_{AB}} = \frac{1}{4 \pi \sqrt{\bar N}} (\lambda R_e)^2 \frac{k_BT \sqrt{\bar N}}{\sigma_{AB} R_e^2}
\end{equation}
measures the strength of fluctuation effects and decreases like $1/\sqrt{\bar N}$. 

At weak segregation $(\lambda R_e)^2 \sim (R_e/\xi)^2 \sim (1-2/\chi N)$ and 
$k_BT \sqrt{\bar N}{\sigma_{AB} R_e^2} \sim (1-2/\chi N)^{-3/2}$ and the capillary parameter increases 
like $\omega \sim [\bar N(1-2\chi N)]^{-1/2}$. In the ultimate vicinity of the critical point the mean 
field theory breaks down and we expect a crossover to a constant value.\cite{OMEGA2}
The divergence of $\omega$ upon approaching the critical point is cut--off around $1-\chi_c N/\chi N 
\sim 1/\bar N$, in accord with the Ginzburg criterium for the crossover from mean field to Ising critical behavior.
In the strong segregation limit the correlation length $\xi$ approaches the 
temperature independent limit $R_e/\sqrt{18}$ and the capillary parameter
scales like $9/[2\pi\sqrt{\chi N/6} \sqrt{\bar{N}}]$. 

Conformational changes are incompletely described by our self-consistent field calculations. 
Within the Gaussian chain model the lateral extension $R_\|$ of a molecule parallel to the surfaces remains always 
unperturbed, i.e., it is independent of the local composition, the distance between the molecules and the surface, 
and the film thickness. Experiments and Monte Carlo simulations, however, do reveal a dependence of the lateral
chain extension on the parameters above. 
Partially, some of these effects can be rationalized as follows: If the film thickness is very thin the lateral
chain extension $R_\|$ has to increase as to restore a constant monomer density. This increase of the lateral 
chain extensions occurs if $\rho R_\|^2 \Delta < N$ or $\Delta/R_e < 1/\sqrt{\bar N}$. Under these conditions 
the chains are quasi two--dimensional and the density of monomers belonging to the same chain inside the coil 
volume is not small. \cite{C1}
An $A$--chain in a $B$--rich environment shrinks as to exchange energetically unfavorable intermolecular contacts
with contacts along the same chain. By reducing its size it increases the density of its own monomers inside the coil volume.
The energy gain upon shrinking is counterbalanced by the loss of conformational entropy. Scaling arguments, Monte Carlo simulations
and SCF calculations\cite{CONF} yield for the relative reduction of a minority chain in the strong segregation limit:
$\Delta R/R \sim \chi N /\sqrt{\bar N}$.

In the limit of infinite interdigitation $\bar N \to \infty$ the above correction to the SCF calculations become small. 
However, even for experimentally relevant chain lengths finite $\bar N$--effects might give rise to sizable corrections
to the SCF calculations (e.g., broadening of the apparent width of $AB$ interfaces by capillary waves).

Besides finite $\bar N$--effects there are other corrections which are not captured by the SCF calculations and which remain important
even in the limit of infinite interdigitation. The finite compressibility of the polymeric fluid, for instance, gives rise to 
packing effects at the surfaces. The monomer density profile in the vicinity of the wall is determined by an intricate interplay between
equation of state effects, loss of conformational entropy, fluid--like packing effects and surface fields. These effects are not included
in the SCF calculations of Gaussian chains, but require a detailed consideration of the molecular architecture and fluid--like packing 
structure. However, we do not expect these effects to change our conclusions qualitatively.

Likewise, it is difficult to find an experimental realization of a symmetric mixture confined into a film with antisymmetric
boundaries. The effects of deviations from perfectly antisymmetric surfaces and the crossover between capillary condensation (for strictly
symmetric boundary fields) to interface localization--delocalization has been explored in the framework of our model.\cite{PRE2} This study shows
that phase diagrams with two critical points also occur for nearly antisymmetric surface fields. The stronger the order of the wetting transition
(the more extended the prewetting line) the more stable is the topology of the phase diagram against small deviations from perfect antisymmetry.

\subsection*{Acknowledgment}
Financial support was provided by the DFG under grant Bi314/17, the VW Stiftung
and PROALAR2000.

\begin{table}[htbp]
\begin{tabular}{|c|c|c|c|c|}
$\Lambda N$ & $\phi_c$ & $\Delta \mu_c/k_BT$ & $T_c=1/\chi_cN$ & $T_t=1/\chi_tN$  \\
\hline
0.5 & 0.2414 & -0.113   &  0.1190   & 0.10791   \\
0.3 & 0.2667 & -0.0574  &  0.2033   & 0.19123   \\
0.2 & 0.3249 & -0.0123  &  0.2770   & 0.27163   \\
0.175 & 0.3584 & -0.00378 &  0.3011   & 0.29867   \\
0.15 & 0.429  & -0.00014 &  0.32949  & 0.32927   \\
\end{tabular}
\caption{
Properties of the antisymmetric system at $\Delta_0=0.9R_e$.
Phase diagrams with two critical points.
} 
\label{tab:resbulk}
\end{table}

\begin{thebibliography}{99}
\bibitem{C1}However, a non--vanishing stiffness of the molecule gives rise to an additional change in the
lateral extension $R_\|$ close to the film surface. This orientation effects is not describable in the Gaussian
chain model and persists also for thick films.
\end{thebibliography}
\end{document}